\documentclass[acmsmall,screen]{acmart}
\usepackage{wrapfig}
\AtBeginDocument{%
  }

\setcopyright{acmlicensed}
\copyrightyear{2025}
\acmYear{2025}
\acmDOI{XXXXXXX.XXXXXXX}

\usepackage{amsmath}            
\usepackage{algorithm}          
\usepackage[noend]{algpseudocode}
\usepackage[normalem]{ulem}
\usepackage{xspace}
\usepackage{makecell}
\newcommand{\mythead}[1]{\thead{\scriptsize #1}}

\usepackage{multirow}
\usepackage{booktabs} 
\usepackage{listings}
\lstdefinestyle{trace}{
  language={},              
  basicstyle=\scriptsize\ttfamily,
  breaklines=true,
  columns=fullflexible,
  frame=single,
  numbers=left,
  numberstyle=\tiny,
  xleftmargin=2em,
  alsoletter=_,               
  classoffset=1,            
  morekeywords={Test_Order_AddItemToCart},   
  keywordstyle=\color{green!60!black}\bfseries,
  classoffset=0 
}
\usepackage[edges]{forest}
\usepackage[table]{xcolor}
\usepackage{siunitx}
\usepackage[most]{tcolorbox}
\usepackage{soul}

\definecolor{lightgray1}{RGB}{225,225,225}
\definecolor{lightgray2}{RGB}{235,235,235}

\newcommand{\toMicroservice}{toMicroservice\xspace}
\newcommand{\approach}{{\textit{Rake}}\xspace}

\newcommand{\mybox}[3]{%
\begin{figure}[htbp]
  \begin{tcolorbox}[
    enhanced,
    arc=8pt,
    colback=white,
    colframe=gray!60,
    boxrule=1pt,
    left=1pt,right=5pt,top=1pt,bottom=1pt,
  ]
    \renewcommand{\do}[1]{%
      \IfBeginWith{##1}{BR}{%
        \colorbox{lightgray2}{\parbox{\dimexpr\linewidth-2\fboxsep}{##1}}%
      }{%
        \colorbox{lightgray2}{\parbox{\dimexpr\linewidth-2\fboxsep}{##1}}%
      }%
      \par\vspace{2pt}%
    }%
    \docsvlist{#3}%
  \end{tcolorbox}
  \caption{#2}\label{#1}
\end{figure}%
}

\newcommand{\mypara}[1]{\par\noindent\textit{#1}}
\definecolor{lightblue}{RGB}{198,219,239}
\definecolor{lightgreen}{RGB}{200,230,200}
\definecolor{lightred}{RGB}{255,200,200}
\newcommand{\best}[1]{\cellcolor{lightblue}{#1}}
\newcommand{\bestother}[1]{\cellcolor{lightred}{#1}}
\newcommand{\bestbcp}[1]{\cellcolor{lightgreen}{#1}}

\begin{document}

\newcommand{\wk}[1]{{\textcolor{red}{~[~\textbf{WK}: \textit{#1} ]}}}
\newcommand{\fabiha}[1]{{\textcolor{gray}{~[~\textbf{TF}: \textit{#1} ]}}}
\newcommand{\saad}[1]{{\textcolor{brown}{~[~\textbf{Saad}: \textit{#1} ]}}}
\newcommand{\neno}[1]{{\textcolor{orange}{~[~\textbf{Neno}: \textit{#1} ]}}}

\title[Identifying Appropriately-Sized Services with Deep Reinforcement Learning]{Identifying Appropriately-Sized Services \\ with Deep Reinforcement Learning}


\author{Syeda Tasnim Fabiha}
\email{fabiha@usc.edu}
\orcid{0009-0006-6821-332X}
\affiliation{%
  \institution{University of Southern California}
  \city{Los Angeles}
  \state{California}
  \country{USA}
}
\author{Saad Shafiq}
\affiliation{%
  \institution{University of Southern California}
  \city{Los Angeles}
  \state{California}
  \country{USA}
}
\email{sshafiq@usc.edu}

\author{Wesley Klewerton Guez Assunção}
\affiliation{%
  \institution{North Carolina State University}
  \city{Raliegh}
  \state{North Carolina}
  \country{USA}}
\email{wguezas@ncsu.edu}

\author{Nenad Medvidović}
\affiliation{%
  \institution{University of Southern California}
  \city{Los Angeles}
  \state{California}
  \country{USA}}
\email{neno@usc.edu}


\begin{abstract}
Service-based architecture (SBA) has recently gained attention in both industry and academia, primarily as a way to modernize legacy systems.
It refers to a design style that enables the development of systems as a suite of small, loosely coupled, and autonomous components (``services'') that encapsulate system functionalities and communicate with each other using language-agnostic APIs.
However, defining appropriately-sized services that encapsulate meaningful, cohesive subsets of system functionality remains a challenge. Additionally, existing work has tended to rely on the availability of certain documentation, access to project personnel (e.g., architects), and \emph{a priori} knowledge of the number of services into which the system should be decomposed. These assumptions do not hold in many real-world scenarios.
Our work aims to address these limitations by employing a deep reinforcement learning-based approach to identify the number of appropriately-sized services from an existing system's implementation artifacts. We present \approach,  a novel reinforcement learning–based approach that leverages a combination of available system documentation and source code to guide service decomposition at the level of implementation methods. \approach does not mandate the presence of specific documentation or the availability of project personnel, and is language-agnostic. Furthermore, \approach features a customizable objective function, enabling decomposition strategies that balance modularization quality and business capability alignment  (i.e., the degree to which a service adequately covers the targeted business capability). 
We applied \approach to four open-source legacy projects and compared it against two state-of-the-art techniques.
On average, \textit{Rake} achieved 7-14\% higher modularization quality and 18-22\% stronger business capability alignment compared to the two existing techniques. Our empirical analysis further reveals that optimizing solely for business context can degrade decomposition quality in tightly coupled systems, underscoring the need for balanced objectives.
\end{abstract}



\received{20 February 2007}
\received[revised]{12 March 2009}
\received[accepted]{5 June 2009}

\maketitle


\section{Introduction}
\label{sec:intro}


Many legacy software systems experience problems including technical debt, architectural drift, and modular degradation, impacting their scalability, maintenance, evolution, and deployment performance (e.g.,~\cite{khadka2014,Tizzei2017, Luz2018}). One emerging solution has been to modernize these systems by decomposing them into service-based architectures (SBA), which comprise clearly delineated, independent services that are easier to manage, scale, and deploy~\cite{Abgaz2023, Wolfart2021, Assuncao2025}, and where
%
%
%
each service is responsible for performing a set of specific business use cases \cite{dragoni2017microservices, Tuli2014,Wang2020}. 
%
%
%
However, a prevalent challenge for this modernization is that the decomposition of legacy code into coherent services that are highly cohesive, loosely coupled, and at the right granularity remains tedious, complex, and error-prone~\cite{Abgaz2023, Wolfart2021}. 

A large amount of existing work has targeted the challenge of identifying services in legacy systems~\cite{Martinez2025}. The resulting approaches have spanned identification of service candidates by analyzing
a system's application domain~\cite{krause2020},  static analysis of the source code~\cite{al2021microservice, nitin2022cargo}, dynamic execution traces capturing runtime dependencies~\cite{jin2021service,toMicroservice2021}, and version-related information (e.g., evolutionary coupling)~\cite{jin2018functionality}. In addition, studies have explored the use of optimization and machine learning (ML) techniques~\cite{desai2021graph, mosaic23}. 
%
%
However, despite their contributions, these studies exhibit key limitations that hinder their applicability in real-world modernization efforts. For example, static and dynamic analysis techniques fail to consider alignment with an application's business capabilities, which is a key concern in  practice~\cite{kalia2021mono2micro}. Moreover, current approaches provide service decompositions without considering appropriately sized services based on business needs~\cite{jamshidi2018microservices, Abgaz2023}. 
Many existing solutions require knowing up-front the number of services into which the system is to be decomposed~\cite{jin2021service, toMicroservice2021, kalia2021mono2micro}, which is not a reasonable assumption~\cite{oumoussa2024evolution}. Furthermore, most studies focus on interface/class-level decomposition~\cite{jin2021service, sellami2022, kalia2021mono2micro, sellami2025rldec},  overlooking the potential for finer, method-level granularity, which can provide actionable insights for system refactoring~\cite{jamshidi2018microservices, wang2024microservice}.
Some studies have explored method-level decomposition~\cite{nitin2022cargo,toMicroservice2021}, but they still rely on predefining the number of services, which limits their adaptability. 

%


Motivated by the promising results of using ML for identifying microservices~\cite{Dehghani2022}, we address the key limitations of prior techniques by combining \emph{(i)}~dynamic analysis driven by a system's business capabilities with \emph{(ii)}~a reinforcement learning (RL) algorithm tailored to the service decomposition problem \emph{(iii)}~at method-level granularity.
We select RL  because it is well-suited to problems where the optimal solution 
is not known beforehand and must be discovered through incremental adaptation to the environment. 
Our approach, named \approach, considers the decomposition problem as an incremental learning problem, where the decomposition is refined at each step based on the system's current state, gradually adapting the optimal number and size of services. This enables our technique to adjust to a legacy system’s unique structure and dependencies without requiring prior knowledge of the target number of services.
%
%
%

\approach operates in two main phases. In the first phase, we combine a systematic identification of business capabilities with the collection of dynamic execution traces aligned with them, allowing us to capture the runtime interaction reflecting the true functional behavior of the system within its business context.
We then transform these traces into a call graph, which holds key information about method-level dependencies, essential for accurately grouping the methods into coherent services. 
This information is fed into an RL environment, where we utilize a \approach-specific adaptation of the Proximal Policy Optimization (PPO) algorithm~\cite{schulman2017proximal,2022ppoeffectiveness}. This algorithm dynamically learns optimal method-to-service assignments through the maximization of a customized objective function designed to satisfy user-specified objectives. 
%
Moreover, by operating at the method level, our approach captures finer-grained structural dependencies that indicate actionable refactoring suggestions essential during the migration, which are often overlooked by class-level approaches.


We have applied \approach to four open-source legacy projects. To evaluate \approach's effectiveness, we compared it against an existing state-of-the-art RL-based technique, RLDec~\cite {sellami2025rldec}, and a method-level decomposition approach, \toMicroservice~\cite{toMicroservice2021} that has been shown to perform well in recent studies~\cite{wang2024microservice}. 
We define four metrics for use in our comparison. Our results show that, overall, \approach outperforms both RLDec and \toMicroservice. At the same time, we identify a small number of individual instances where that has not been the case and discuss the reasons behind \approach's relative underperformance.
Our empirical analysis also reveals that optimizing solely for the business context often leads to poor decomposition quality, especially in systems with tightly coupled components. We offer a set of possible solutions, implemented in \approach, for mitigating such scenarios.

The key contributions of this paper are as follows:
\begin{itemize}
    \item \approach, a novel semi-automated RL-based 
    approach to identify appropriately sized services from legacy projects at method-level granularity, allowing a customizable objective function tailored to user-specific needs.

    \item In-depth analysis of trade-offs between business capability alignment and modularization quality, especially in tightly coupled systems.
    
    \item An extensive empirical evaluation on four legacy projects, using four metrics, and comparing against two state-of-the-art approaches, along with the provision of a replication package.\footnote{All artifacts of \approach are made available at: \url{https://doi.org/10.5281/zenodo.17102986}}  
\end{itemize}

\looseness-1
The paper is organized as follows. Section 2 details our approach, and Section 3 details our evaluation strategy. Section 4 presents the evaluation results and discusses their implications. Section 5 summarizes the related work, and Section 6 discusses the validity threats. Finally, Section 7 presents our conclusions.

\section{\approach: A Semi-automated RL-based Approach}
\label{sec:approach}

\begin{figure*}
\centering
\includegraphics[width=1\textwidth]{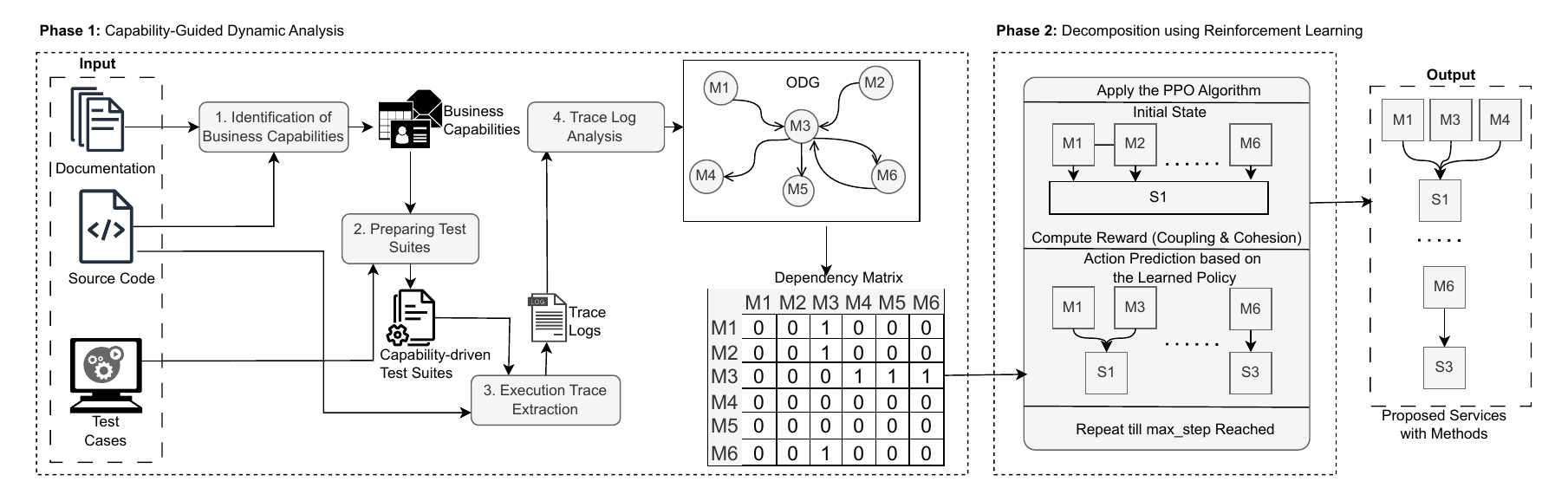}
\vspace{-7mm}
\caption{Overview of \textit{Rake}}
\vspace{-4mm}


\label{fig:approach}
\end{figure*}



In this paper, we introduce a new deep RL-based approach, \textit{Rake}, that provides method-level decomposition of a legacy software system into services.
%
%
To achieve this, \textit{Rake} incrementally analyzes the system and reorganizes its methods into appropriately sized services that are aligned with the system's implemented business capabilities,\footnote{A business capability is a set of related business use cases that are responsible for serving one main responsibility~\cite{dragoni2017microservices}.} guided by an objective function. Our approach considers the system’s available artifacts—documentation, source code, and test cases—and produces service candidates at the method level.
Determining effective service boundaries in legacy systems is challenging because the artifacts are often incomplete or inconsistently aligned with the system’s business logic. For example, test suites may fail to cover all functionalities~\cite{regressionTest2018}, and source code may not always follow naming conventions that reflect business intent~\cite{mcburney2016empirical}. Moreover, while some systems adopt Domain-Driven Design (DDD) practices that conveniently align the codebase with business goals~\cite{Silva2019, fan2017migrating}, many legacy systems do not, making capability identification more difficult. Prior approaches often assume completeness of test cases or direct alignment of source code with system objectives~\cite{jin2018functionality}. \textit{Rake} explicitly avoids such assumptions by validating, supplementing, and cross-checking artifacts to ensure broad coverage of business capabilities.

A high-level overview of \approach is depicted in Figure~\ref{fig:approach}.
The main inputs to \textit{Rake} are a combination of the system's available artifacts: documentation, source code, and test cases. 
From these inputs, we extract the system’s business capabilities, map them to executable test suites, 
and collect execution traces aligned with those suites. Where gaps between the available test cases and capabilities are identified, \textit{Rake} generates additional test cases by navigating through the system’s user interface (UI), ensuring that the dynamic behavior of all capabilities is exercised. The collected execution traces are then analyzed to produce a dependency matrix that captures method-to-method interactions in the system. This matrix provides the foundation for RL–based decomposition and ensures that the analysis reflects the actual runtime behavior of the system.

We assume that the system undergoing decomposition must include some form of documentation from which the high-level business goals can be identified.
This may take the form of requirements specifications, technical documents, UML diagrams, user manuals, README files, or descriptions of the system’s core objectives in online resources. 
Preferably, the source code will follow an established
naming convention that reflects the high-level business logic, but \approach does not require this. In cases where such conventions are
not present, we rely on the business logic previously identified from the documentation. Another source of information that could further ameliorate the possible shortcomings in a system's documentation and codebase is the personnel working on the system (e.g., its architects)~\cite{toMicroservice2021,fan2017migrating}. We do not assume the availability of project staff, but \textit{Rake} would naturally include the information they provide.


\textit{Rake}’s analysis proceeds in two phases (see Figure~\ref{fig:approach}). In the first phase—\textit{Capability-Guided Dynamic Analysis}—\textit{Rake} {systematically identifies business capabilities from available artifacts,} prepares executable test suites mapped to each capability for dynamic analysis, collects resulting execution traces of the system functionalities, and analyzes the obtained logs. The output of this phase is a system-wide dependency matrix reflecting method-to-method call relationships. 
In the second phase—\textit{Decomposition Using Reinforcement Learning}—\textit{Rake} leverages the dependency matrix to construct a reinforcement learning (RL) agent that iteratively assigns methods to services. To this end, we adapt the Proximal Policy Optimization (PPO) algorithm to the decomposition problem, providing a new formulation of the state space, action space, and customizable reward function that effectively capture the system's internal structure and dependencies. Notably, \approach does not assume \emph{a priori} knowledge of the number of services.
Instead, \approach's RL agent always aims to generate decompositions that align closely with a system's business capabilities while maintaining the service decomposition quality.
The final output of \approach is a set of method-level service candidates, optimized through iterative interaction between the agent and the environment.

It is worth noting that the evaluation subjects used in our study to date are Java-based systems, which is why the first phase of \approach's current implementation (detailed below)
employs Java-specific tools. However, the approach itself is language-agnostic and can be applied to any programming language, provided that the required dependency information can be extracted in a similar manner. 

Next, we introduce a running example that will help to illustrate \textit{Rake}'s details. The subsequent subsections elaborate on \textit{Rake}'s two analysis phases.

\subsection{Example System} 
\begin{figure*}[t!]
\centering
\includegraphics[width=0.9\textwidth]{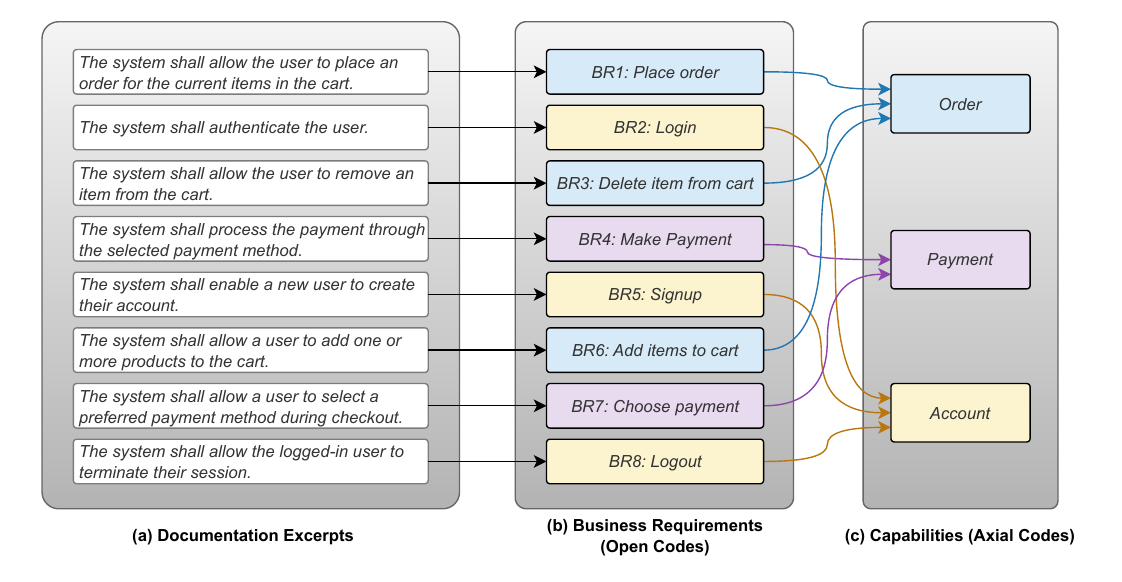}
\vspace{-5mm}
\caption{Scattered requirements grouped into categories through qualitative coding 
}
\label{fig:br_to_cap}
\vspace{-4mm}
\end{figure*}

We use an e-commerce system  to illustrate how \textit{Rake} operates.
Figure~\ref{fig:br_to_cap}(a) shows a set of business requirements extracted from the available documentation.
From these requirements, we can infer that the system supports basic functions such as user registration, order placement, and making payments, although the documentation does not explicitly define the system's business capabilities. 
The system's code structure suggests the presence of high-level domains such as \textit{AccountService}, \textit{UserService}, and \textit{OrderService}, as shown in Figure~\ref{fig:legacy_struct}.
A closer inspection of the code reveals significant architectural erosion, such as functionality that should belong to a single concern being scattered across multiple implementation classes. For example, in the \textit{OrderService} class shown in Figure~\ref{fig:cross_deps}, the \texttt{placeOrder} method is not limited to order processing. Instead, it first checks user eligibility through the \texttt{isEligible} method, and then invokes the \textit{PaymentService} to charge the user before saving the order. The former should be a concern either for the \textit{Account} or \textit{User} business logic, and the latter shows direct reliance on another service. 
This tight coupling scatters responsibilities across services, introduces hidden dependencies, and makes the order processing logic rely directly on account validation and payment execution.
Such dependencies make it difficult to modify individual features or to scale components independently.
\textit{Rake} aims to reorganize the system at the method level, disentangling business logic and clustering methods into cohesive services. 












\begin{figure}[b!]
  \centering
  \begin{minipage}[t]{0.32\linewidth}
    \begin{tcolorbox}[colback=white, colframe=black!40,
                      boxrule=0.5pt, valign=top,
                      height=0.260\textheight] 
      \begin{forest}
        for tree={
          font=\footnotesize,
          grow'=0,
          child anchor=west,
          parent anchor=south,
          anchor=west,
          calign=first,
          inner ysep=-2pt,
          outer ysep=-1pt,
          forked edges,
          edge path={
            \noexpand\path [draw, \forestoption{edge}]
            (!u.south west) +(4pt,0) |- (.child anchor) \forestoption{edge label};
          },
          before typesetting nodes={ if n=1 {insert before={[,phantom]}} {} },
          fit=band,
          before computing xy={l=8pt},
        }  
        [<e-commerce\_system>
          [src/main/java/org
            [web ]
            [model ]
            [service
              [AccountService.java ]
              [UserService.java ]
              [OrderService.java ]
              [PaymentService.java ]
            ]
            [test
              [AccountServiceTest.java ]
              [OrderServiceTest.java ]
              [PaymentServiceTest.java ]
            ]
          ]
          [webapp ]
          [target]
        ]
      \end{forest}
    \end{tcolorbox}
    \vspace{-4mm}
    \caption{Code structure for the \\ e-commerce system.}
    \label{fig:legacy_struct}
  \end{minipage}\hfill
  \begin{minipage}[t]{0.67\linewidth}
    \begin{tcolorbox}[colback=white, colframe=black!40,
                      boxrule=0.5pt, valign=top,
                      height=0.260\textheight] 
\begin{lstlisting}[
  language=Java,
  basicstyle=\ttfamily\tiny,
  keywordstyle=\color{blue}\bfseries,
  stringstyle=\color{brown},
  commentstyle=\color{gray}\itshape,
  numbers=left, numberstyle=\tiny\color{gray},
  frame=none, breaklines=true, showstringspaces=false
]
@Service
public class OrderService {
  private final OrderActionBean order;
  private final PaymentService ps; 
    private boolean isEligible(User user) {
   // returns if user is eligible or not
  }
  public Order placeOrder(User user, Cart cart) {
    if (!isEligible(user)) { 
      throw new UnauthorizedException();
    }
    ps.charge(cart.total(), userId);
    return OrderActionBean.save(new Order(user.getId(), cart));
  }
}
\end{lstlisting}
    \end{tcolorbox}
    \vspace{-4mm}
    \caption{Cross-concern dependencies inside \textit{OrderService.java}.}
    \label{fig:cross_deps}
  \end{minipage}
  \vspace{-4mm}
\end{figure}


\subsection{Phase 1: Capability-Guided Dynamic Analysis}


As outlined above, this phase focuses on capturing method-level interactions that are relevant to a system's business logic. As with prior work (e.g., ~\cite{kalia2021mono2micro}), the goal is to exclude unnecessary methods (e.g., utility and helper) and produce cohesive services. 
To capture the pertinent method-level interactions, in this phase, we perform dynamic analysis guided by the business capabilities, in four steps, illustrated in Figure~\ref{fig:approach}: \textit{(i)}~identify the business capabilities and related use cases for the system under study, \textit{(ii)}~prepare test suites based on these capabilities, \textit{(iii)}~extract the dynamic execution traces by executing the test suites, and \textit{(iv)}~analyze these traces through source code representations such as call graph and dependency matrix. We next elaborate on each step. 

\subsubsection{Systematic Identification of Business Capabilities and Use Cases} 
\label{sub_section:identify_bp}
Knowing the system's business capabilities is critical because the decomposed services should align with these capabilities~\cite{dragoni2017microservices, Dragoni2017}. 
%
A number of existing approaches assume that capabilities are readily available, rely on domain experts, or leverage pre-existing architectural knowledge~\cite{Silva2019, fan2017migrating, toMicroservice2021}. However, these information sources are frequently not available in practice. 
Another line of existing work does analyze system artifacts~\cite{jin2018functionality,kalia2021mono2micro,jin2021service}, but without a targeted procedure for deriving business capabilities. In contrast, \approach provides a systematic approach that integrates multiple artifacts to identify capabilities using a qualitative coding procedure~\cite{seaman1999qualitative} from grounded theory methods~\cite{corbin2014basics}.

First, in the exploratory step, we conduct a manual review of available system documentation and supplementary artifacts, including requirements specifications, technical documents, UML diagrams, user manuals, and/or README files. During this review, we look for \emph{business requirements} (BRs)---recurring functional entities and requirements that reflect high-level business logic. We analyze the identified BRs using open coding~\cite{corbin2014basics}: we assign codes to individual statements or descriptions reflecting the system's functionalities, and refine and expand the set of codes as new BRs emerge throughout the review. We treat the BRs yielded by this process as the system's business use cases. Figure~\ref{fig:br_to_cap}(b) illustrates the open codes (i.e., business requirements) derived from the requirements in the sample system documentation presented in Figure~\ref{fig:br_to_cap}(a). 

\begin{wraptable}{r}{0.55\textwidth}
\vspace{-2mm}
\centering
\scriptsize
\caption{Business Capabilities and Use Cases for the\\ e-commerce system
\vspace{-2mm}}
\label{tab:use_case}
\begin{tabular}{p{0.925cm} p{0.65cm} p{1.85cm} p{2.7cm}}
\toprule
\textbf{Capability} & \textbf{Corresp. BR} & \textbf{Use Case Name} & \textbf{Use Case Description} \\
\midrule
\multirow{3}{*}{Account} & BR2 & \textit{Login}    & Login to the system             \\
                         & BR8 & \textit{Logout}   & Logout from the system          \\
                         & BR5 & \textit{Signup}   & Register a new user account     \\
\midrule
\multirow{3}{*}{Order}   & BR6 & \textit{AddItemsToCart}     & Add items to shopping cart      \\
                         & BR3 & \textit{DeleteItemFromCart} & Delete item from cart           \\
                         & BR1 & \textit{PlaceOrder}         & Place an order                  \\
\midrule
\multirow{2}{*}{Payment} & BR7 & \textit{ChoosePayment}      & Choose payment method           \\
                         & BR4 & \textit{MakePayment}        & Make payment                    \\
\bottomrule
\end{tabular}
\end{wraptable}

Next, we use axial coding~\cite{corbin2014basics}, where we systematically examine relationships among the open codes (i.e., BRs), grouping them into higher-level categories that represent coherent business capabilities, such as \textit{Account}, \textit{Order}, and \textit{Payment} shown in Figure~\ref{fig:br_to_cap}(c). During this step, we integrate complementary evidence from the system's source code. \approach analyzes the source code using a static parser that extracts structural elements of the system, such as classes, methods, and their signatures. 
In principle, any parser capable of producing this information can be used; in our implementation, we rely on the JavaParser~\cite{java_parser}
framework. We extend this parser with a keyword-based search utility to identify classes consisting of the following suffixes and annotations, which adhere to established naming conventions outlining high-level business logic or functional responsibilities~\cite{Silva2019, Wolfart2021, Francesco2018industrial, evans2004domain}: \texttt{controller}, \texttt{service}, \texttt{repositories}, \texttt{entities}, and \texttt{domain}.
In our running example, the search returned matches for the \texttt{service} keyword, with the corresponding classes located in the \textit{service} and \textit{test} folders, as shown in Figure~\ref{fig:legacy_struct}.
We link these findings to the initial codes from the documentation, consolidating synonymous or overlapping meanings into unified categories and refining them into stable business capabilities. For instance, classes such as \textit{AccountService} and \textit{UserService} were mapped to the \textit{Account} categories identified during axial coding. Figure~\ref{fig:br_to_cap}(c) depicts the final categories derived from the open codes for our example e-commerce system. Finally, we consolidate the categories into a set of business capabilities that are supported across the different artifacts.
%
Table~\ref{tab:use_case} illustrates the identified capabilities and their corresponding use cases.

To ensure the accuracy of the identified business capabilities, in the empirical evaluation reported in this paper, the first two authors independently completed the qualitative coding and reviewed all mappings between codes and capability categories.
Any discrepancies in interpretation or assignment were resolved through structured discussions until consensus was reached. It is important to note that, in practice, the requirements-to-capabilities assignment may be subjective, hampered by insufficient information, missing artifacts, or a lack of clarity. In such cases, the resolution step may not be possible. \approach mitigates this by allowing stakeholders to assess multiple interpretations in an exploratory fashion, by processing each of the multiple interpretations in the manner elaborated in the remainder of this section, and comparing the resultant service decomposition quality metrics as elaborated in Section~\ref{sec:evaluaion}.

\subsubsection{Preparing Test Suites} \approach requires dynamic analysis of the system to uncover the method interactions relevant to the identified business capabilities. This process involves preparing focused test suites that exercise the relevant parts of the  code~\cite{toMicroservice2021}.
We design executable tests corresponding to each use case identified during qualitative coding  (recall Table~\ref{tab:use_case}). We first attempt to reuse the system's existing functional tests (e.g., unit and usage-based test cases) and map them to the corresponding business use cases.
If the  available test cases are
not relevant to the system's business use cases (or are missing altogether), we create additional test cases by navigating the system's user interface, aiming to have at least one test case for each business use case.
The set of test cases corresponding to a business capability forms a test suite that is used during the dynamic analysis.

To facilitate traceability, we annotate each test case with its capability and use-case identifier, allowing us to map the execution traces back to their originating capabilities. In our e-commerce system, the capability \textit{Order} is associated with the \textit{AddItemsToCart}, \textit{DeleteItemFromCart}, and \textit{PlaceOrder} use cases, which are implemented as executable test cases grouped into the \textit{Test\_Order} test suite. Then, for example, the test case corresponding to the use case \textit{AddItemToCart} is annotated as \texttt{Test\_Order\_AddItemsToCart} (highlighted in green in the execution trace Listing~\ref{lst:traces}); \texttt{Test\_Order} denotes the suite corresponding to the \textit{Order} capability and \texttt{AddItemsToCart} specifies the executable test case.

\subsubsection{Execution Trace Extraction} 
Completing the dynamic analysis involves running the generated test suites and collecting the resulting execution traces. 
To facilitate that, we instrument the subject system's source code using probes from the Kieker~\cite{kiekerv15}
 open-source framework for application monitoring. Kieker relies on AspectJ, a widely used Aspect-Oriented Programming (AOP) extension, which we employ to capture all method execution records triggered by incoming UI requests~\cite{kiczales1997aspect, gradecki2003mastering}. 
This process follows established instrumentation guidelines~\cite{van2012kieker,hasselbring2020kieker}, with entry and exit points for methods explicitly defined, indicating where Kieker's monitoring probes are to be inserted. These probes are responsible for recording and logging the method executions. 
AspectJ allows weaving the probes automatically into the defined points (e.g., all source classes), eliminating the need to manually modify individual methods. 
 After completing the instrumentation, the system is deployed, and 
 the functional test suites for each business capability are executed using Katalon Studio,\footnote{Katalon Studio is an automated testing IDE built upon Selenium that supports creating and executing tests on the web browser. See: \url{https://docs.katalon.com/katalon-studio/about-katalon-studio}} a UI test automation tool. 
 The corresponding trace log files are generated in this process. The log files allow us to analyze and visualize traces of the method executions per business capability. 

{
\begin{lstlisting}[style=trace,
caption={Excerpt of collected execution trace for \texttt{Test\_Order\_AddItemToCart}}\\,
label={lst:traces}
]
$2;1753777000000000001;Test_Order_AddItemToCart
$1;1753777000000001201;OrderActionBean.addItemToCart(java.lang.String,int);<no-session-id>;2499076000000000001;1753777000000001101;1753777000000001400;localhost;2;1
$1;1753777000000002202;AccountService.getCartByUser(java.lang.String);<no-session-id>;2499076000000000001;1753777000000001210;1753777000000001290;localhost;3;2
$1;1753777000000003203;OrderService.addItemToCart(model.Cart,jva.lang.String,int);<no-session-id>;2499076000000000001;1753777000000001230;1753777000000001275;localhost;4;3
$1;1753777000000004204;OrderActionBean.setQuantity(int);<no-session-id>;2499076000000000001;1753777000000001500;1753777000000001550;localhost;5;2
\end{lstlisting}
}

Listing~\ref{lst:traces} shows an excerpt of the collected execution traces for the e-commerce system's test case \texttt{Test\_Order\_AddItemToCart}. 
The log file comprises two types of records: \textit{(i)}~trace-header records (\texttt{\$2}, Line 1 in Listing~\ref{lst:traces}) mark the start of the execution trace for a given test case; and \textit{(ii)}~operation-execution records (\texttt{\$1}, Lines 2-5 in Listing~\ref{lst:traces}) {capture the individual method invocations associated with the test case, annotated with attributes indicating the order of these} invocations~\cite{van2009continuous}.

\subsubsection{Trace Log Analysis}
\label{subsec:trace_log}
\begin{wrapfigure}{r}{0.4\textwidth} 
    \vspace{-5mm}
    \centering
    \includegraphics[width=0.35\textwidth]{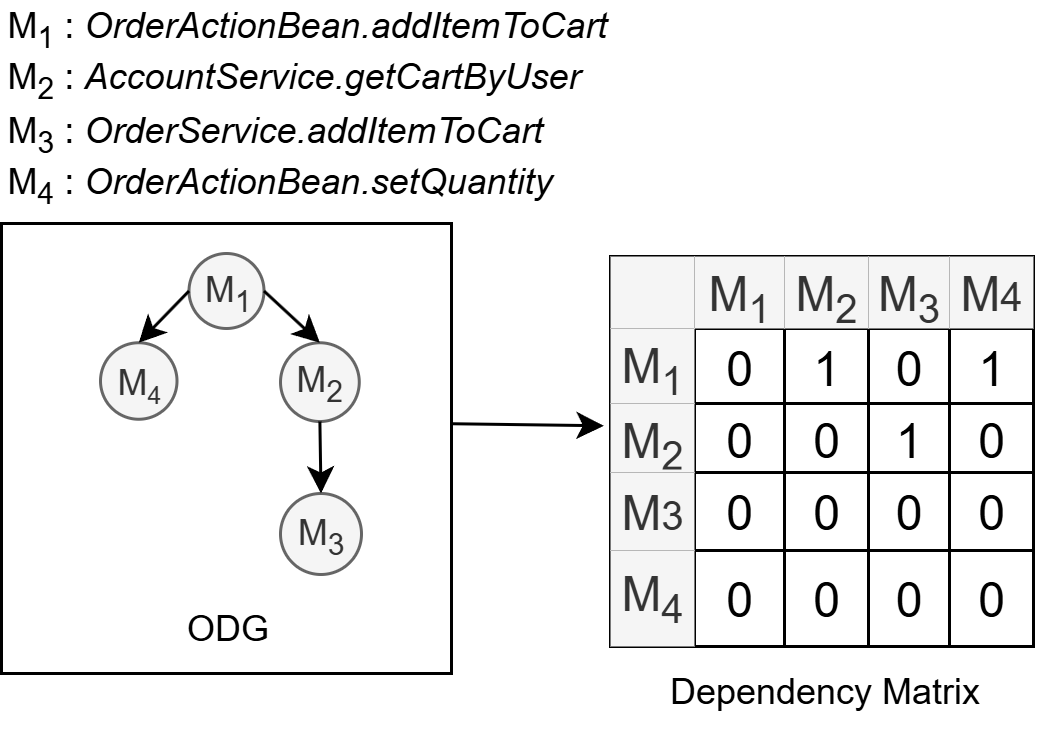}
    \vspace{-5mm}
    \caption{ODG for trace log from Listing~\ref{lst:traces}}
    \label{fig:odg}
\end{wrapfigure}

\approach uses the built-in trace log analysis feature from Kieker~\cite{kieker_trace}
which generates directed Operation Dependency Graphs (ODGs)~\cite{kieker_codebase}
from the trace logs for each identified capability. The ODGs contain method caller and callee information. Figure~\ref{fig:odg} presents a simplified ODG for the execution log from Listing~\ref{lst:traces}, where each node represents a method and the edges capture the call relationships. Since the example e-commerce system has three business capabilities, \approach generates three corresponding ODGs that hold information about the methods associated with each capability. This method-to-capability mapping is essential for evaluating the decomposed service-to-capability alignment.
Finally, we merge all of the ODGs into a system-wide dependency matrix, while consolidating the common nodes and their interrelationships. The system-wide matrix is used as input to the service decomposition process elaborated next.


\subsection{Phase 2: Decomposition using Reinforcement Learning}
\label{sub_sec:phase2}

In the second phase, \approach uses the call-dependency matrix obtained from the first phase to construct an RL agent for automated service decomposition. The agent’s objective is to learn to produce high-quality decompositions by iteratively assigning methods to services. To achieve this, \approach employs a neural network–based policy model leveraging the widely used PPO algorithm~\cite{schulman2017proximal,2022ppoeffectiveness}.
The key contribution of \approach lies in the formalization of this learning process, realized through six RL components: \textit{observation}, \textit{state}, \textit{action}, \textit{reward}, \textit{episode}, and \textit{environment}. 

\begin{figure}
\centering
\includegraphics[width=0.95\textwidth]{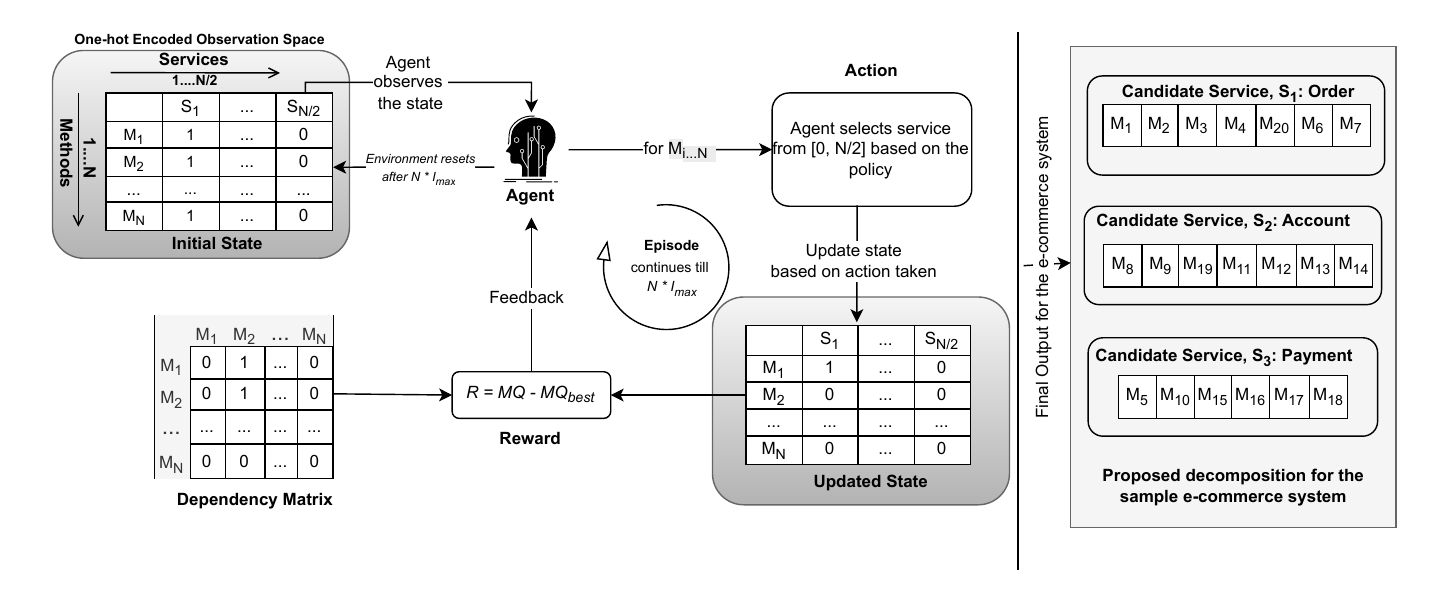}
\Description{Overview of \approach's second phase in the context of service decomposition.}
\vspace{-4mm}
\caption{Overview of \approach's second phase [proposed decomposition on the right]}
\label{fig:ppo}
\vspace{-4mm}
\end{figure}

Figure~\ref{fig:ppo} illustrates how these components interact throughout the training process, as well as the final decomposition produced by the trained agent for the example e-commerce system (depicted in the right portion of the figure). 
At the start of training \approach's RL agent \emph{observes} the initial \emph{state} of the \emph{environment}, which corresponds to a trivial decomposition where all methods derived from the dependency matrix are assigned to a single service. Based on this state, the agent selects an \emph{action} (obtained from a policy network) that causes a change in state and receives feedback in the shape of a \emph{reward}, an indication of whether the selected action resulted in improved decomposition (in this case, modularization quality derived from the dependency matrix generated in \approach's first phase). The environment then transitions to the updated state, and the process continues until the terminal condition is reached. This entire cycle is called an \emph{episode}. The agent’s behavior is incrementally refined at each episode, guided by the reward feedback. At the end, the trained agent provides the candidate decomposition based on the policy it has learned. Next, we detail how these RL components are formalized.

\textbf{Observation.}
\looseness-1 In RL, the observation space defines what information about the environment is available to the agent at each step of the learning process.
\approach constructs the environment to correspond to the system under decomposition, and the observation space is configured to capture how methods are assigned to candidate services at any given time.
A key design decision in this phase is to 
constrain the maximum number of services to at most half of the total number of methods $N$, a property 
suggested by the existing clustering problem~\cite{scanniello2010using}. This bound reduces the dimensionality of the observation space while still maintaining sufficient flexibility to explore meaningful decompositions. 
Formally, given $N$ nodes from the ODG, the observation space is represented as a one-hot encoded $N{\times}\frac{N}{2}$ matrix, where each row corresponds to a method and each column denotes a service (as highlighted in the upper-left segment of Figure~\ref{fig:ppo}). An entry of $1$ at $(i,j)$ indicates that method $M_i$ is currently assigned to service $S_j$. The PPO agent interacts with this space across multiple {episodes}, where in each episode it iterates over the methods, assigns them to services, receives feedback in the form of rewards, and progressively improves its decomposition strategy.

\textbf{State.} 
In RL, the state represents the environment at a given time and provides the basis for the agent’s decision-making. In \approach's setting, the state corresponds directly to the observation, since the agent has full visibility of the current method-to-service assignments.
We design the state in the same one-hot encoded form as the observation space. The initial state assigns all methods to a single service, providing the starting point for exploration (note the leftmost column of the matrix in the upper-left of Figure~\ref{fig:ppo} populated with ``1'' values). As training progresses, the state is updated to reflect the agent’s current decomposition choices, with each reassignment of a method altering the matrix accordingly (center-right segment of Figure~\ref{fig:ppo}). This update process continues throughout learning, enabling the PPO agent to iteratively refine the decomposition strategy.



\textbf{Action.}  
\approach designs the action space to correspond to a method-to-service assignment during the decomposition process. At each step, the agent iterates through the methods and selects a service for the current method under consideration. 
The policy network, representing the agent’s decision-making strategy, outputs a probability distribution over the available action space, and one action is sampled from this distribution at each step. The quality of the undertaken actions is guided by the selection of the reward function, which we describe next.
Thus, we design the action space to be discrete, ranging from $[0, (\frac{N}{2}-1)]$, where each value denotes the service identifier to which the method is assigned. 
This design choice yields a linear action space, which contributes to the scalability of our approach: the agent explores decompositions method-by-method, rather than considering exponentially many possible services 
simultaneously.

\textbf{Reward.}
The central element of the RL algorithm is the reward $(R)$ objective, which provides feedback to the agent after each action and drives the learning process toward desirable outcomes.
Designing an appropriate reward is crucial, since it determines how the agent evaluates its progress and what patterns it learns to reinforce.
In \approach, this means that the reward must capture the quality of a decomposition in a way that aligns with established principles of good modularization. 
To quantify decomposition quality, we rely on widely adopted modularization metrics,
 \textit{Cohesion} $(C_h)$ and \textit{Coupling} $(C_p)$. $C_h$ highlights intra-connectivity, which is the coherent behavior of components internal to a service, while $C_p$ highlights inter-connectivity, which minimizes dependencies across components spanning different services~\cite{newman2021building}. 
$C_h$ and $C_p$ serve as building blocks of the widely used \textit{Modularization Quality (MQ)} metric, which determines the quality of decomposed services~\cite{mancoridis1998using} and is the primary objective of \approach. 
A higher \textit{MQ} value signifies better decomposition quality. We formally define $C_h$, $C_p$, \textit{MQ}, and \textit{R} next.



\textit{Cohesion $(C_h)$} --  
For each service $S_i$,  cohesion is calculated using the intra-connectivity defined by Mancoridis~\cite{mancoridis1998using}. It measures the proportion of actual internal dependencies (number of unique method-to-method calls) relative to the maximum possible internal dependencies (all possible method-to-method calls) in a service:  
$
    {C_h} = \frac{1}{k} \sum_{i=1}^{k} {C_h}_{(S_i)}, \quad {C_h}_{(S_i)} = \frac{\mu_i}{N_i^2}
$,
where $N_i$ is the number of methods in  $S_i$, and $\mu_i$ is the number of intra-service (internal unique method-call) dependencies among those methods. The overall cohesion of the system is the average across all $k$ services.

\textit{Coupling $(C_p)$} --  
Coupling between two services $S_i$ and $S_j$ is defined as the inter-connectivity between them, normalized by the maximum possible number of cross-service dependencies:  
$ 
{C_p} = \frac{2}{k(k-1)} \sum_{i,j=1}^{k}  {C_p}_{(S_i,S_j)}, \quad  {C_p}_{(S_i,S_j)} = \frac{\nu_{i,j}}{2 \cdot N_i \cdot N_j}
\label{eq:cp}
$,
where $\nu_{i,j}$ is the number of unique method calls between $S_i$ and $S_j$, and $N_i$, $N_j$ are the total numbers of methods in each service. The coupling of the entire system is measured by the average across all pairs of the $k$ services, where $k>1$.

\textit{Modularization Quality $(MQ)$} -- 
This metric captures the trade-off between cohesion and coupling:  
\begin{equation}
    \label{eq:mq}
    MQ =
    \begin{cases}
        {C_h} - {C_p}, & \text{if } k > 1 \\[6pt]
        {C_h}_{(S_1)}, & \text{if } k = 1
    \end{cases}
\end{equation}
The $MQ$ score, ranging from $-1$ to $1$, increases when services are internally cohesive and externally decoupled, thereby encouraging well-modularized decompositions.

\textit{Reward Function $(R)$} --  
The reward is designed to incentivize the agent toward decompositions with higher modularization quality. At each step, we compute $MQ$ for the action taken by the agent on the current decomposition. The \emph{step reward} is defined as the difference between the current $MQ$ and the best $MQ$ observed so far within the episode.
\begin{equation}
    \label{eq:reward}
    R = MQ - MQ_{\text{best}}
\end{equation}
If the current $MQ$ exceeds $MQ_{\text{best}}$ recorded in the episode, both $MQ_{\text{best}}$ and the corresponding clustering state configuration are updated. 
If the reward is neutral or negative, it penalizes the agent since the action did not improve the decomposition. This design of  \emph{R} 
ultimately guides the decomposition toward higher-quality modular boundaries.

While \approach employs $\textit{MQ}$ as the default reward objective, it is designed to be customizable and to allow alternate objectives. To demonstrate this flexibility, we employed an alternate objective that measures the capability alignment of the decomposed services. This objective and its use in \approach is detailed in Section$~\ref{sec:evaluaion}$, and the corresponding results are reported in Section$~\ref{sec:result}$.


\textbf{Episode.} 
In RL, an episode is defined as a complete sequence of interactions between the agent and the environment, beginning with an initial state and ending upon reaching a terminal condition (see Figure~\ref{fig:ppo}).
In \approach, the agent interacts with the environment across multiple episodes, progressively improving its decomposition based on the feedback received.
The initial configuration and the terminal condition of episodes are explicitly defined to suit the decomposition task in \approach's design.
Each episode begins with all methods initialized into a single service and proceeds by reassigning methods to services based on the agent’s actions. 
A full traversal of all $N$ methods constitutes one \textit{pass ($P$)}. After the last method is processed, the environment either \emph{(i)} resets the current node to the first method and starts a new iteration, or \emph{(ii)} resets the environment if the maximum number of passes $P_{\max}$ has been reached. 
Therefore, the episode terminates when either all methods have been processed for the maximum number of iterations $P_{\max}$, or when the maximum number of timesteps $T = N \times P_{\max}$ is reached.
The introduction of $P_{\max}$ ensures that the episode length remains bounded, preventing unproductive looping and keeping training computationally tractable.
As the agent iterates over methods, during each iteration, the environment evaluates the quality of the decomposition measured by $\textit{MQ}$ and $MQ_{\text{best}}$. At termination, the environment returns the best decomposition discovered during the episode. 
This design allows the agent to refine service assignments iteratively within the same episode before it enters into next episode, consistent with RL practices~\cite{beck2025tutorial}. 
Recalling our e-commerce system example, suppose the ODG yields a total of $20$ methods, and the maximum passes per episode is set to $P_{\max}=3$. In this case, the episode length would be $T = 20 \times 3 = 60$.

\textbf{Environment.} 
The environment manages the complete iterative cycle by encapsulating the \textit{state}, \textit{action}, \textit{reward}, and \textit{episode} progression. In our design, the environment handles the method-to-service assignment process and evaluates decomposition quality at each iteration. At every timestep $t$, given the current state $s_t$, \approach's RL agent performs an action $a_t$ (i.e., assigns the current method to a service), and receives a reward $r_t$, which is based on the resulting modularization quality $MQ$ (recall Equation~\ref{eq:mq}).
Through this interaction, the environment continuously provides feedback to the agent, enabling it to refine its policy over successive timesteps.  
Thus, the agent is trained in an episodic fashion, with each episode refining the decomposition until the maximum number of training steps is reached. The final decomposition is obtained from the trained agent by extracting all non-empty services and their assigned methods from the best-performing configuration. The final output of the algorithm is therefore a set of proposed services, defined at method-level granularity and optimized through iterative interactions between the agent and the environment. The right portion of Figure~\ref{fig:ppo} illustrates the final decomposition of the e-commerce system into three services. 
For readability, we annotated each candidate service with the corresponding capability based on the service-to-capability mapping obtained from \approach's first phase.

\section{Evaluation Strategy}
\label{sec:evaluaion}
We evaluate \approach empirically, on a set of real-world subject systems, using a set of metrics that were adopted from literature and extended to better reflect the nature of the service decomposition problem. 
Our evaluation aims to answer the following research questions.
\begin{itemize}

    \item \textbf{\textit{RQ1}} -- How effective is \approach in producing service decompositions compared to existing baselines?

    \item \textbf{\textit{RQ2}} -- How do different objective functions employed by \approach impact the alignment of system decomposition with business capability?
    
\end{itemize}
The remainder of this section details our evaluation subjects, metrics used to evaluate \approach, the existing baselines against which we compare \approach, and our evaluation setup. 

\subsection{Subject Systems}

We used four Java-based legacy web applications in our evaluation: 
JPetStore~\cite{jpetstore_codebase},
PetClinic~\cite{petclinic_codebase},
Apache Roller~\cite{roller_codebase},
and Social Edition~\cite{edition_codebase}.
Table~\ref{tab:projects} shows their numbers of methods and identified business capabilities, as well as the coverage achieved through dynamic analysis for the identified capabilities.
These systems were selected for several reasons: \emph{(i)}~they are available as open-source, deployable, and runnable; \emph{(ii)} they provide user manuals and/or online documentation~\cite{petclinic_doc,jpetstore_doc,edition_doc,roller_doc},
which was helpful in  identifying their respective business capabilities;  \emph{(iii)}~they vary in complexity, ranging from PetClinic, a simple management system for a pet clinic with straightforward business operations, to Roller, a large-scale multi-user blogging platform with thousands of methods and complex interactions; and \emph{(iv)}~they have been used as benchmarks in prior research on service decomposition~\cite{sellami2025rldec,kalia2021mono2micro,jin2021service,wang2024microservice}.  

\begin{wraptable}{r}{0.4\textwidth}
\centering
  \caption{Evaluation subjects and their dynamic analysis coverage.}
  \label{tab:projects}
  \scriptsize 
  \renewcommand{\arraystretch}{1.05} 
  \begin{tabular}{lccc}
    \toprule
    \textbf{System} & \textbf{Methods} & \textbf{Capab.} & \textbf{Cover. (\%)} \\
    \midrule
    PetClinic      & 160   & 4 & 114 (71\%)  \\
    JPetStore      & 290   & 3 & 227 (78\%)  \\
    Social Edition & 645   & 5 & 92 (14\%)   \\
    Apache Roller  & 4607  & 5 & 1308 (28\%) \\
    \bottomrule
  \end{tabular}
\end{wraptable}

In general, a software system may rely on facilities such as event/message passing, database interactions, network communication, file I/O, etc. At application-level, such functionalities are typically realized via method invocation-based APIs~\cite{laliwala2008event}. For instance, the entity-relation model, widely adopted in databases~\cite{millett2015patterns}, is typically implemented in entity/data model classes and their constituent methods. Similarly, many event/message-based solutions (e.g., Android) rely on mechanisms such as callbacks that are ultimately implemented using methods. Thus, our dynamic analysis focuses on application-layer methods associated with business capabilities.

As shown in Table~\ref{tab:projects}, the proportion of methods that are aligned with business capabilities, and thus identified by \approach during dynamic analysis, ranged from $14\%$ in Social Edition to $78\%$ in JPetStore. Note that larger systems contain a relatively lower proportion of methods dedicated to implementing business capabilities compared to the two smaller systems. While fully uncovering the underlying reasons behind this would require a more in-depth study that is beyond our work reported in this paper, the observed trend was unsurprising in that we anticipate larger systems to have a lot more internal, utility functionality that is not directly related to business use cases.





\subsection{Metrics}
To evaluate the quality of the decompositions produced by \approach, we select four widely adopted metrics that capture different aspects of modularization quality~\cite{sellami2025rldec,toMicroservice2021, jin2021service, kalia2021mono2micro}.
These include business capability alignment, cohesion–coupling balance, inter-service dependency, and interface exposure. Together, the metrics provide a comprehensive basis for assessing the effectiveness of our approach.

\subsubsection{Business Capability Alignment}
We place particular emphasis on the extent to which \approach's produced services 
preserve capability boundaries in accordance with the single-responsibility principle~\cite{martin2014srp}. 
Prior work has proposed two entropy-based metrics for this purpose: \textit{Business Context Purity (\textit{BCP})}~\cite{kalia2020mono2micro} and \textit{Domain Independence (\textit{DI})}~\cite{wang2024microservice}.


{\textit{BCP}} measures how consistently business use cases are distributed within each service, i.e., whether methods associated with the same use case are grouped together without unnecessary dispersion across multiple services.
For each service, \textit{BCP} calculates the entropy of the distribution of methods across business use cases. When most methods in a service belong to the same use case, entropy is low and {\textit{BCP}} is high. Since in our setting we have a direct mapping of methods to business use cases and subsequently to business capabilities, we calculate this entropy at the capability level. The {\textit{BCP}} score is then normalized across all services,
where higher values indicate stronger alignment between services and individual capabilities~\cite{wang2024microservice}.  
For a given service $i$, let $n_b$ be the number of methods associated with capability $b$, and $n$ the total number of methods in $i$. 
Then, the probability of a method from  service $i$ appearing in capability $b$ is defined as is: $P(b) = \tfrac{n_b}{n}$, and the entropy of service $i$ is: 
$H(i)~=~-~\sum_b P(b)\log P(b)$. 
We define \textit{BCP} as: $1 - \tfrac{1}{M}\sum_{i=1}^M H(i)$, where $M$ is the number of services in the decomposition, scaled to $[0,100]$. 
To illustrate using our running e-commerce example, $\approach$ provides a decomposition with three candidate services, $S_1$, $S_2$, and $S_3$, as shown in Figure~\ref{fig:ppo}. Each candidate service contains all methods that belong to a single, unique business capability. The method-to-capability mapping obtained from the trace log analysis (recall Section~\ref{subsec:trace_log}) is $S_1$: \textit{Order}, $S_2$: \textit{Account}, and $S_3$: \textit{Payment}. The respective probability distributions for the three services are $[\frac{7}{7},0,0]$, $[0,\frac{7}{7},0]$, and $[0,0,\frac{6}{6}]$, yielding entropy $H(i) = 0$ for each service. As a result, the value of $\textit{BCP}$ is the maximum $100\%$.

We note that {\textit{BCP}} has an important limitation: It only measures \textit{intra-service affinity}, i.e., whether a service spans multiple capabilities. \textit{BCP} does not penalize cases where a single capability is scattered across several services. {For instance, if 
services $S_1$ and $S_2$  in the e-commerce example both consist solely of methods that belong to the {\emph{Order}} capability, each $H(i)$ is still $0$ and $\textit{BCP}$ remains $100\%$, even though the $\emph{Order}$ capability is fragmented across multiple services.}
The second entropy-based metric, \textit{DI}, was introduced to address this complementary perspective by measuring dispersion at the capability level~\cite{wang2024microservice}. For each capability, \textit{DI} computes the entropy of the capability's distribution across services. If most methods of a capability are concentrated in a single service, {\textit{DI}} will be high; if a capability is fragmented across many services, {\textit{DI}} decreases. Thus, {\textit{DI}} captures \textit{inter-service affinity}, but it does not penalize cases where multiple capabilities are mixed within the same service.

Recognizing that neither traditional {\textit{BCP}} nor {\textit{DI}} alone fully captures capability alignment, we propose \emph{Adjusted \textit{BCP} (ABCP)}, a new metric that integrates the two. Our metric penalizes both types of misalignment, i.e., multiple capabilities mixed within a service and a single capability scattered across multiple services. We define \textit{ABCP} as a weighted average of \textit{BCP} and \textit{DI}:
\begin{equation}
    \label{eq:true_BCP}
    \text{\textit{ABCP}} = 0.5 \times \text{\textit{BCP}} ~+~ 0.5 \times \text{\textit{DI}}
\end{equation}
The weighted average ensures that intra-service and inter-service entropy are treated with the same level of importance. It also means that sacrificing either dimension lowers the overall score, preventing one form of alignment from masking deficiencies in the other.
\textit{ABCP} therefore provides a more comprehensive measure of business capability alignment, balancing internal service cohesion with overall capability coherence across the decomposition.

\subsubsection{Modularization Quality (MQ)}
As discussed earlier in the reward component of \approach (recall Section~\ref{sub_sec:phase2}), \textit{MQ} captures decomposition quality in terms of cohesion and coupling~\cite{mancoridis1998using}, 
where higher values indicate higher-quality decompositions. Its definition is provided in Equation~\ref{eq:mq}.

\subsubsection{Inter-service Call Percentage (ICP)}
\textit{ICP} measures the proportion of runtime calls that cross service boundaries, thereby reflecting inter-service dependencies~\cite{kalia2020mono2micro}. For a pair of services, the call ratio can be expressed in a log-scaled form to reduce the dominance of disproportionately large call counts~\cite{sellami2025rldec}:
$
ICP = \frac{\sum_{i=1}^{K} \sum_{\substack{j=1 \\ j \neq i}}^{K} icp_{i,j}}
           {\sum_{i=1}^{K} \sum_{j=1}^{K} icp_{i,j}}, \quad icp_{i,j} = \sum_{m_k \in i} \sum_{m_l \in j} \big(\log(inv(m_k, m_l)) + 1\big)
$, where $K$ is the total number of services, $m_k \in i$ and $m_l \in j$ denote methods in services $i$ and $j$, respectively, and $inv(m_k, m_l)$ is the number of runtime calls between them. 
Lower \textit{ICP} values are desirable as they indicate fewer cross-service dependencies. 

\subsubsection{Interface Number (IFN)}
\textit{IFN} captures the number of published interfaces of a service, reflecting the extent to which the service exposes functionality externally~\cite{jin2021service}.
\textit{IFN} can be adapted to the method-level decomposition supported by \approach, by interpreting interfaces as externally visible methods of each service. Formally,  
$
IFN = \frac{1}{N} \sum_{j=1}^{N} ifn_j, \quad ifn_j = |I_j|
\label{eq:ifn}
$, where $I_j$ denotes the set of methods in service $j$ that are invoked from other services, and $N$ is the number of services in a system. Lower \textit{IFN} values indicate that services expose fewer entry points, suggesting a cleaner separation of responsibilities.
We include \textit{IFN} as one of our evaluation criteria because minimizing the number of externally visible methods helps to ensure that a service has focused responsibility. 

\subsection{Baselines}
To provide an objective evaluation of \approach, we compare it against two state-of-the-art service decomposition techniques that represent the closest alternatives available: \textit{(i)}~{RLDec}~\cite{sellami2025rldec}, the only reinforcement learning-based approach proposed to date, which operates at the class level; and \textit{(ii)}~{\toMicroservice}~\cite{toMicroservice2021}, a method-level decomposition approach that combines static and dynamic analysis and has been reported to outperform other existing method-level techniques~\cite{wang2024microservice}. Since both of these techniques generate multiple service decompositions with the goal of enhancing structural quality, given the same primary objective of $\approach$, we select only those decompositions that yield the highest MQ within each technique.

\subsubsection{RLDec}
This approach employs a Deep Q-Learning agent to explore different ways of grouping classes into candidate services. It relies solely on static analysis, constructing structural and semantic feature matrices from source code artifacts, which are then used as input to guide the agent through reward functions based on decomposition quality. RLDec introduces two formulations, Sequential and Flattened, as well as combined variants that leverage both structural and semantic features. RLDec's evaluation showed that it outperforms clustering-based and evolutionary approaches in terms of structural and semantic cohesion. Note that a reproducible package was not available for end-to-end execution and analysis, however, the class-level services decompositions generated for all the projects employed in our study were available. To enable a fair comparison with our method-level approach, we converted their class-level decompositions into method-level representations. {Specifically, for each service, we prepared a script to automatically extract the methods that belong to all the classes within that service, effectively yielding method-level service decompositions.}
Among these converted decompositions, we selected the one with the highest \textit{MQ} score to represent RLDec in our comparative analysis. 

\subsubsection{\toMicroservice}
This approach combines static and dynamic analysis to construct a graph-based representation of methods and their interactions, and then applies a multi-objective search-based optimization using NSGA-III~\cite{nsga-iii}.
The optimization simultaneously considers four criteria: coupling, cohesion, network overhead, and feature modularization. In the context of \toMicroservice, feature modularization is based on use cases serving as entry points, which requires input from domain experts~\cite{Carvalho2024}. 
A reproducible package was made available by the authors, which we employed to re-run the approach on our subject applications. 
Since \toMicroservice requires the number of services to be specified in advance, their original study reported results based on a setup involving 5, 7, and 10 services. Although, and as discussed previously, an important feature of \approach is that it does not require that the number of services be specified in advance, for comparison purposes we followed the same setup as \toMicroservice for each of our subject systems. 

\subsection{Experimental Setup}
\approach conducts the capability-guided dynamic analysis on each subject system independently. Afterwards, it trains the RL agent for each system using the derived dependency matrix.
Each system is trained for 1,500 episodes. This number was determined empirically: preliminary experiments showed that training beyond 1,500 episodes did not yield significant improvements in quality, while substantially increasing training cost. Since the length of an episode is determined by the number of methods in the system's dependency matrix (with a minimum of three passes over all methods), the total number of training steps varied across systems. In smaller systems, episodes contained fewer steps, whereas larger systems had longer episodes and therefore consumed more training steps overall. After training, we evaluate the learned policy by selecting the decomposition with the highest quality score under the optimization objective. 
%
%
Recall from Section~\ref{sub_sec:phase2} that \approach's optimization function is customizable, allowing the integration of any desired objective in principle. To demonstrate this, we integrate \textit{ABCP} as an alternate objective into \approach. 
%
Thus, we experiment with two optimization objective configurations in the reward component: one, based on \textit{MQ},  encourages structural modularization quality, while the other, based on  \textit{ABCP}, emphasizes alignment with business capabilities. 


\section{Results and Discussion}
\label{sec:result}

\looseness=-1
In this section, we present our empirical findings. 
Specifically, we evaluate a total of four service-decomposition tool implementations. The first three tools were introduced above: \emph{(i)}~\approach(\textit{MQ}), the default setting of \approach optimized for \textit{MQ}, as detailed in Section~\ref{sec:approach}; and \emph{(ii)}~RLDec and \emph{(iii)}~\toMicroservice, the state-of-the-art baselines introduced in Section~\ref{sec:evaluaion}. In addition, to evaluate \approach's ability to incorporate alternate optimization objective functions, we also include \emph{(iv)} \approach(\textit{ABCP}), an implementation of \approach optimized for the \emph{ABCP} metric introduced in Section~\ref{sec:evaluaion}.

Each tool implementation is assessed using the four metrics that were introduced previously. Two of the metrics are also used as \approach's optimization criteria: \textit{MQ} reflects structural modularity, while \textit{ABCP} reflects business capability alignment. The remaining two metrics capture finer-grained characteristics of a decomposition:  \textit{ICP}  measures the proportion of runtime calls that reflect inter-service dependencies, while \textit{IFN} quantifies the number of published interfaces of a service thus capturing the extent of external functionality exposure.
Using \textit{MQ} and \textit{ABCP}, both, as objective functions that guide how services are formed and as evaluation metrics allows us to assess the extent to which \approach is able to optimize to a target objective and to subsequently evaluate the resulting services for additional properties of interest. 

Our evaluation was framed by the two research questions introduced in Section~\ref{sec:evaluaion}. To answer \emph{RQ1}, we evaluate the default \approach \textit{(MQ)} against the state-of-the-art baselines RLDec and \toMicroservice across the four metrics. To answer  \emph{RQ2}, we evaluate \approach \textit{(ABCP)} against the baselines as well as against \approach~\textit{(MQ)}. 
Table \ref{tab:result} reports the values of the four metrics corresponding to the four tool implementations across the four subject systems. For each metric, the cells highlighted in \best{blue} represent the instances in which \approach \textit{(MQ)} outperformed or matched the 
two baselines
for each of the four subjects. The cells highlighted in \bestbcp{green} indicate the instances in which \approach \textit{(ABCP)} matched or outperformed the remaining three tools for each of the four subjects. Finally, the cells highlighted in \bestother{red} indicate instances in which the default \approach \textit{(MQ)} was outperformed by one of the baseline tools, even if \approach \textit{(ABCP)} performed better than the baselines for a given subject. The table allows us to examine how each \approach optimization strategy performs under its own objective and how well it generalizes to the other quality metrics.




\subsection{\emph{RQ1} -- Effectiveness in Producing Service Decompositions} 
The goal of this \emph{RQ} is to evaluate whether \approach can generate decompositions of higher structural quality than the two baseline techniques. Specifically, we examine how effectively \approach improves cohesion and reduces coupling (\emph{MQ}). Additionally, we measure the extent to which \approach maintains a reasonable degree of alignment with business capabilities (\emph{ABCP}), while minimizing the number of calls crossing service boundaries (\textit{ICP}) and number of exposed interfaces (\textit{IFN}).

Perhaps unsurprisingly, \approach \textit{(MQ)}, the default settings of \approach where the reward objective is \textit{MQ}, achieves the highest \textit{MQ} values across all four systems as compared to the two baseline techniques:
on average, \approach yields a $7\%$ higher value of \textit{MQ} compared to RLDec and $14\%$ higher than \toMicroservice. 
\approach \textit{(MQ)} also outperforms the baselines across the other three metrics on average. In terms of \textit{ABCP}, \approach \textit{(MQ)}  outperforms RLDec by $18\%$  and \toMicroservice by $22\%$ across all four subject systems.

\begin{figure}[b]
\vspace{-4mm}
\noindent \begin{minipage}[c]{0.6\textwidth}
\centering
  \setlength{\tabcolsep}{7pt}
  \scriptsize
 \captionof{table}{Evaluation metrics obtained for the four techniques across the subject systems (higher is better for MQ and BCP; lower is better for ICP and IFN).}
 \renewcommand{\tabcolsep}{4pt}
\vspace{-2mm}
  \begin{tabular}{llccccc} 
    \toprule
    \textbf{System} & \textbf{Technique} & \textbf{\textit{MQ}}$^{\uparrow}$ & \textbf{\textit{ABCP}}$^{\uparrow}$ & \textbf{\textit{ICP}}$^{\downarrow}$ & \textbf{\textit{IFN}$^{\downarrow}$} & \textbf{\# Services} \\
    \midrule
    \multirow{4}{4em}{{PetClinic}} &  \approach (\textit{MQ})     & \best{0.127} & 71.725 & \best{0.009} & \best{0.500} & 2 \\
    &\approach (\textit{ABCP})    & 0.059 & \bestbcp{84.190} & 0.162 & 1.300 & 6 \\
    & RLDec             & 0.083 & \bestother{77.22} & 0.045 & 1.667 & 6 \\
    &\toMicroservice    & 0.041 & 42.195 & 0.453 & 5.200 & 5 \\
    \midrule
    \multirow{4}{4em}{{JPetStore}} &    \approach (\textit{MQ})               & \best{0.169} & \best{79.310} & 0.266 & 22.667 & 3 \\
    &\approach  (\textit{ABCP})              & 0.104 & \bestbcp{80.200} & 0.373 & 13.200 & 5 \\
    &RLDec                       & 0.026 & 32.955 & \bestother{0.056} & \bestother{11.500} & 2 \\
    &\toMicroservice              & 0.062 & 56.885 & 0.352 & 20 & 5 \\
    \midrule
    \multirow{4}{4em}{{Social Edition}} &    \approach (\textit{MQ})                & \best{0.172} & \best{79.275} & \best{0} & \best{0} & 3 \\
    &\approach (\textit{ABCP})               & 0.132 & \bestbcp{81.265} & \bestbcp{0} & \bestbcp{0} & 2 \\
    &RLDec                        & 0.163 & 62.86 & 0.138 & 1.330 & 3 \\
    &\toMicroservice               & 0.013 & 46.47 & 0.227 & 0.500 & 10 \\
    \midrule
    \multirow{4}{4em}{{Apache Roller}} &    \approach (\textit{MQ})               & \best{0.199} & \best{49.53} & 0.054 & \best{27.33 }& 3 \\
    &\approach (\textit{ABCP})              & 0.001 & \bestbcp{77.625} & \bestbcp{0} & \bestbcp{0} & 2\\
    &RLDec                       & 0.120 & 36.645 & \bestother{0.035} & 69.000 & 2 \\
    &\toMicroservice              & 0.010 & 47.06 & 0.143 & 37.40 & 5 \\
    \bottomrule
  \end{tabular}
  \label{tab:result}
\end{minipage}
\hskip 8pt
\begin{minipage}[c]{0.35\textwidth}
 \centering
  \vspace{-2mm}
  \captionof{table}{Degree of method overlap among business capabilities identified by \approach.}
\vspace{-2mm}
  \scriptsize 
  \renewcommand{\tabcolsep}{3pt}
  \begin{tabular}{lccc}
   
  \toprule
  \\[-4.25mm]
  \textbf{System} & \mythead{\textbf{Shared}\\ \scriptsize \textbf{Methods}} & \mythead{\scriptsize \textbf{Total}\\ \scriptsize \textbf{Methods}} & \mythead{\textbf{Overlap} \\ \textbf{(\%)}} \\
  \\[-4.25mm]
    \midrule
    PetClinic      & 21  & 114  & 18.4\% \\
    JPetStore      & 9   & 227  & 3.96\% \\
    Social Edition & 9   & 92   & 9.78\% \\
    Apache Roller  & 140 & 1308 & 10.70\% \\
    \bottomrule
      &   &   & \\
        &   &   & \\
        &   &   & \\
        &   &   & \\
        &   &   & \\
        &   &   & \\
        &   &   & \\
        &   &   & \\        
        &   &   & \\
        &   &   & \\
             &   &   & \\
   &   &   & \\ \end{tabular}
  \label{tab:overlap}
\end{minipage}
\end{figure}

However, PetClinic presents an exception: \approach \textit{(MQ)}'s \emph{ABCP} value ($71.73\%$) is lower than RLDec's ($77.22\%$). 
A closer inspection of the code reveals that PetClinic has a relatively high degree of method overlap among its business capabilities. We quantify this overlap as the ratio of methods shared across two or more business capabilities to the total number of methods obtained from the dynamic analysis.
Table~\ref{tab:overlap} presents this information for each subject system. As seen in the table, PetClinic exhibits significantly higher method overlap ($18.4\%$) than the other three subjects, 
indicating that PetClinic’s business logic is implemented in a highly interdependent manner. As a result, while \approach \textit{(MQ)} produces solutions with strong modularization quality, the limitations inherent in the code introduce a trade-off that reduces the achievable \textit{ABCP}. 


{A somewhat related limitation is observed in the context of Apache Roller, the largest system in our study. \approach~\textit{(MQ)} outperforms the two baselines in terms of \textit{ABCP}, demonstrating that it can scale while preserving structural quality and capability alignment. However, the \textit{ABCP} value it yields ($49.53\%$) is markedly lower than for the other three subject systems.
We hypothesize that this is caused at least in part by Apache Roller's decomposition characteristics. While Apache Roller's method overlap is lower than that of PetClinic and comparable to Social Edition, Apache Roller is a significantly larger system and thus has a substantially higher actual number of shared methods (140, as shown in Table~\ref{tab:overlap}) relative to the number of capabilities (5, as shown in Table~\ref{tab:projects}), which in turn negatively influences \textit{ABCP}. Additional study is required to further confirm this hypothesis.}





\approach~(\textit{MQ}) also improves the \textit{ICP} and \textit{IFN} metrics for PetClinic and Social Edition. This indicates that 
maximizing \textit{MQ} can simultaneously reduce structural complexity in smaller, tightly coupled systems. However, RLDec yields lower (i.e., better) \textit{ICP} values with its JPetStore and Apache Roller service decompositions. This suggests that RLDec favors decompositions with fewer inter-service dependencies, and it appears to accomplish that by yielding small numbers of services: RLDec produces only two services for each of the two subjects, which naturally minimizes inter-service coupling and interfaces. However, this comes at the cost of reduced modularization quality and weaker business alignment. For example, in the case of JPetStore, the value of \textit{MQ} yielded by RLDec is 85\% lower than \approach~\textit{(MQ)} and 79\% lower than \approach~\textit{(ABCP)}. Similarly, the value of \textit{ABCP} yielded by RLDec is nearly 60\% lower than both implementations of \approach.  

\toMicroservice consistently underperforms across all metrics, with notably low \textit{MQ} and \textit{ABCP} scores. The relatively high (i.e., worse) \textit{ICP} and \textit{IFN} values it yields further indicate that the resulting decompositions are fragmented and poorly aligned to business capabilities. The reason lies in the design of \toMicroservice: it optimizes decompositions based on use cases rather than business capabilities. While a set of related use cases typically corresponds to a single capability, optimizing directly at the use-case level induces decompositions that are too fine-grained. As a result, methods that should belong to the same capability are often split across services.


Beyond comparing against baselines, another benefit of our default setting \approach~\textit{(MQ)} is that it allows architects to cross-check business capability interpretations, a task usually undertaken before the migration. Given the same structural decomposition, two different interpretations of business capabilities may yield very different \textit{ABCP} values. By comparing these values, \approach \textit{(MQ)} provides a practical signal to practitioners about which interpretation is more consistent with the system’s underlying structure. This feature of \approach is particularly useful in large legacy systems (such as Roller), where capability boundaries are often ambiguous, and it supports experts in validating or refining their capability prior to migration.

\subsection{\emph{RQ2} -- Impact of Different Objective Functions} 
The goal of this \emph{RQ} is to examine how the choice of reward function in \approach influences the resulting decompositions. Specifically, we contrast decompositions generated when optimizing for \textit{MQ} against those optimizing for \textit{ABCP}, and analyze the trade-offs between structural quality and business capability alignment. 

As shown in Table~\ref{tab:result}, \approach \textit{(ABCP)} achieves the highest \textit{ABCP} values across all systems. Especially notable is the extent to which this setting improves upon our \approach \textit{(MQ)} default in the context of the largest subject system, Apache Roller: \approach \textit{(MQ)} already yielded the value of \textit{ABCP} that was higher than the two baselines, but \approach \textit{(ABCP)} improved upon that by another 36\%. 
This demonstrates that \approach optimizes its decompositions effectively when business capability alignment is strictly prioritized.
However, this comes at the expense of structural quality. Nearly across the board, \approach~\textit{(ABCP)}'s performance is worse than that of \approach \textit{(MQ)} for the remaining metrics. This is especially egregious in the case of the \textit{MQ} value for Apache Roller: \approach \textit{(ABCP)} produces a decomposition whose modularization quality is 95\% worse than \approach \textit{(MQ)}'s. 

The above example illustrates how capability-driven optimization can fragment very large systems: optimizing solely on business concerns can improve business alignment, but leads to weakened modular structure and highly coupled services. Rather than a weakness, this reflects an intended feature of \approach. Namely, \approach's customizable objective function allows its users to optimize based on their preference:
when structural cohesion and reduced coupling are the priority, practitioners can choose to optimize for \textit{MQ}; on the other hand, they can optimize for \textit{ABCP} when alignment with business capabilities is more critical. This flexibility allows \approach to support different migration contexts, giving architects explicit control over the inherent trade-offs. Moreover, \approach's reward can be defined as a combination, by assigning desired weights to \textit{MQ} and \textit{ABCP} to obtain decompositions that balance modularization quality and business alignment in a desired manner. 

\section{Related Work}
\label{sec:related}
The goal of software modernization is to create organized software systems that are more modular, easier to develop, and maintain~\cite{toMicroservice2021}. To this end, practitioners have increasingly been converting their existing systems to service-based architectures (SBAs)~\cite{taibi2017processes,Abgaz2023}. However, recent studies have shown that migrating legacy code to SBAs without understanding the business capabilities may result in increased complexity, higher costs, and suboptimal performance \cite{su2023back,su2024modular}. 
Several approaches have explored clustering algorithms to group system entities based on various criteria. Anup et al.~\cite{kalia2021mono2micro} utilize dynamic analysis followed by a hierarchical clustering approach to group classes into services. Another approach involves using a neural model for static code representation and then applying hierarchical clustering to identify service candidates \cite{al2021microservice}. Some studies have employed coupling-based graph clustering techniques to solve the same problem \cite{mazlami2017extraction, gysel2016service}. However, most of the existing clustering approaches provides service suggestions at class-level granularity, overlooking the need for finer-grained decomposition at the level of methods. 

Recent studies have also focused on search-based techniques to solve software modularization problems \cite{carvalho2020performance, khoshnevis2023search, sellami2022, saidani2019towards}. Most of the search-based approaches employ evolutionary genetic algorithms to propose service candidates based on execution traces extracted from a target system~\cite{jin2018functionality, jin2021service, saidani2019towards, carvalho2020performance}. These 
techniques commonly combine clustering and genetic algorithms (e.g.,~NSGA-II) to identify functional units and group them into services \cite{jin2018functionality, jin2021service}. Some of the existing studies propose their solutions based strictly on structural dependencies in the source code \cite{saidani2019towards}, while others consider both structural and semantic dependencies \cite{sellami2022}. Researchers have also explored objective criteria other than coupling and cohesion for decomposing  software~\cite{carvalho2020performance}. However, the existing approaches have largely left unaddressed the challenge of finding an optimal number of services for a legacy system while performing service decomposition at method-level granularity. Moreover, none of these approaches explicitly addresses the challenge of identifying business capabilities, which is a critical step for ensuring meaningful business-aligned decompositions.
\vspace{-.5pt}
\section{Threats to Validity}
\mypara{External Validity}. 
A common potential threat is the generalizability of the results reported in a study such as ours. While we employ projects implemented in Java, we designed \approach to be language-agnostic. The implementation of our custom keyword-based parser needs to be extended to support other languages, but that extension is reasonably straightforward.
\vspace{1.5mm}

\mypara{Internal Validity}.
Another potential threat is related to our selection of subject systems. We mitigated this threat by including widely-adopted projects spanning a wide range of sizes. Another concern is the review process to identify the business capabilities. We followed well-established guidelines~\cite{corbin2014basics} for qualitative analysis to systematically extract the high-level business functionalities. We further ensured the reliability of this assessment through independent validation from two of the authors.
\vspace{1.5mm}

\mypara{Construct Validity}.
This threat refers to how well the conducted experiments confirm the measured outcomes. We address this concern by including the standard metrics employed in literature~\cite{kalia2021mono2micro, jin2021service, sellami2025rldec} for each decomposition generated by \approach. We further provide a trade-off analysis of how two key metrics (\textit{MQ} and \textit{BCP}) influence the reward function, and hence the decompositions.
\vspace{1.5mm}

\mypara{Conclusion Validity}.
This threat concerns the authenticity of outcomes in the study and the possibility of a fair comparison with existing baselines. We rigorously followed the implementations to reproduce the competing approaches. Note that RLDec does not have publicly available end-to-end implementation details, but only the resultant class-based decompositions. We systematically converted these decompositions to the method-level and compared with the other techniques in our study using well-defined standard metrics.

\section{Conclusion}

In this paper, we have described \approach, a novel approach based on RL that is aimed at suggesting an appropriate number of appropriately-sized service candidates for a legacy software system at method-level granularity. Our goal is to assist software architects in the re-design and modernization of their systems' architectures. \approach has the potential to make a notable practical improvement to software modularization, by reducing the time developers must spend on coming up with the best way to decompose their software. 

\looseness-1 As part of our planned future work, we aim to explore a range of reward functions in \approach, both, to generate better decomposition quality, and to evaluate \approach's effectiveness and flexibility on a wider variety of legacy systems. Furthermore, we intend to evaluate \approach's usability by conducting user studies with developers and architects. Finally, we will continue to benchmark \approach against the state-of-the-art techniques, to position our work within the broader context of the existing solutions.

\section{Data Availability}
This paper contains data as a replication package. 
All the necessary code, datasets, and instructions as well as other resources required to reproduce the results of this study, are available in an online repository: \url{https://doi.org/10.5281/zenodo.17102986}


\bibliographystyle{ACM-Reference-Format}
\bibliography{references}

@inproceedings{desai2021graph,
  title={Graph neural network to dilute outliers for refactoring monolith application},
  author={Desai, Utkarsh and Bandyopadhyay, Sambaran and Tamilselvam, Srikanth},
  booktitle={AAAI Conference on Artificial Intelligence},
  volume={35},
  pages={72--80},
  year={2021}
}

@article{Assuncao2025,
author = {Assun\c{c}\~{a}o, Wesley K. G. and Marchezan, Luciano and Arkoh, Lawrence and Egyed, Alexander and Ramler, Rudolf},
title = {Contemporary Software Modernization: Strategies, Driving Forces, and Research Opportunities},
year = {2025},
issue_date = {June 2025},
publisher = {ACM},
address = {New York, NY, USA},
volume = {34},
number = {5},
issn = {1049-331X},
doi = {10.1145/3708527},
journal = {ACM Trans. Softw. Eng. Methodol.},
month = may,
articleno = {142},
numpages = {35}
}

@ARTICLE{Carvalho2024,
  author={Carvalho, Luiz and Colanzi, Thelma Elita and Assunção, Wesley K. G. and Garcia, Alessandro and Pereira, Juliana Alves and Kalinowski, Marcos and de Mello, Rafael Maiani and de Lima, Maria Julia and Lucena, Carlos},
  journal={IEEE Transactions on Software Engineering}, 
  title={On the Usefulness of Automatically Generated Microservice Architectures}, 
  year={2024},
  volume={50},
  number={3},
  pages={651-667},
  doi={10.1109/TSE.2024.3361209}
}

@inproceedings{mosaic23,
  title={From monolithic to microservice architecture: an automated approach based on graph clustering and combinatorial optimization},
  author={Filippone, Gianluca and Mehmood, Nadeem Qaisar and Autili, Marco and Rossi, Fabrizio and Tivoli, Massimo},
  booktitle={2023 IEEE 20th International Conference on Software Architecture (ICSA)},
  pages={47--57},
  year={2023},
  organization={IEEE}
}

@inproceedings{dragoni2017microservices,
  title={Microservices: How to make your application scale},
  author={Dragoni, Nicola and Lanese, Ivan and Larsen, Stephan Thordal and Mazzara, Manuel and Mustafin, Ruslan and Safina, Larisa},
  booktitle={International Andrei Ershov Memorial Conference on Perspectives of System Informatics},
  pages={95--104},
  year={2017},
  organization={Springer}
}

@inproceedings{nitin2022cargo,
  title={Cargo: Ai-guided dependency analysis for migrating monolithic applications to microservices architecture},
  author={Nitin, Vikram and Asthana, Shubhi and Ray, Baishakhi and Krishna, Rahul},
  booktitle={Proceedings of the 37th IEEE/ACM International Conference on Automated Software Engineering},
  pages={1--12},
  year={2022}
}

@article{jamshidi2018microservices,
  title={Microservices: The journey so far and challenges ahead},
  author={Jamshidi, Pooyan and Pahl, Claus and Mendon{\c{c}}a, Nabor C and Lewis, James and Tilkov, Stefan},
  journal={IEEE Software},
  volume={35},
  number={3},
  pages={24--35},
  year={2018},
  publisher={IEEE}
}

@inproceedings{mazlami2017extraction,
  title={Extraction of microservices from monolithic software architectures},
  author={Mazlami, Genc and Cito, J{\"u}rgen and Leitner, Philipp},
  booktitle={2017 IEEE Int. Conference on Web Services (ICWS)},
  pages={524--531},
  year={2017},
  organization={IEEE}
}

@inproceedings{gysel2016service,
  title={Service cutter: A systematic approach to service decomposition},
  author={Gysel, Michael and K{\"o}lbener, Lukas and Giersche, Wolfgang and Zimmermann, Olaf},
  booktitle={Service-Oriented and Cloud Computing: 5th IFIP WG 2.14 European Conference, ESOCC 2016, Vienna, Austria, September 5-7, 2016, Proceedings 5},
  pages={185--200},
  year={2016},
  organization={Springer}
}

@article{su2023back,
  title={Back to the Future: From Microservice to Monolith},
  author={Su, Ruoyu and Li, Xiaozhou and Taibi, Davide},
  journal={arXiv preprint arXiv:2308.15281},
  year={2023}
}

@inproceedings{su2024modular,
author = {Su, Ruoyu and Li, Xiaozhou},
title = {Modular Monolith: Is This the Trend in Software Architecture?},
year = {2024},
isbn = {9798400705601},
publisher = {Association for Computing Machinery},
address = {New York, NY, USA},
url = {https://doi-org.libproxy2.usc.edu/10.1145/3643657.3643911},
doi = {10.1145/3643657.3643911},
booktitle = {Proceedings of the 1st International Workshop on New Trends in Software Architecture},
pages = {10–13},
numpages = {4},
location = {Lisbon, Portugal},
series = {SATrends '24}
}

@article{taibi2017processes,
  title={Processes, motivations, and issues for migrating to microservices architectures: An empirical investigation},
  author={Taibi, Davide and Lenarduzzi, Valentina and Pahl, Claus},
  journal={IEEE Cloud Computing},
  volume={4},
  number={5},
  pages={22--32},
  year={2017},
  publisher={IEEE}
}

@article{khoshnevis2023search,
  title={A search-based identification of variable microservices for enterprise SaaS},
  author={Khoshnevis, Sedigheh},
  journal={Frontiers of Computer Science},
  volume={17},
  number={3},
  pages={173208},
  year={2023},
  publisher={Springer}
}

@inproceedings{carvalho2020performance,
  title={On the performance and adoption of search-based microservice identification with tomicroservices},
  author={Carvalho, Luiz and Garcia, Alessandro and Colanzi, Thelma Elita and Assun{\c{c}}{\~a}o, Wesley KG and Pereira, Juliana Alves and Fonseca, Baldoino and Ribeiro, M{\'a}rcio and de Lima, Maria Julia and Lucena, Carlos},
  booktitle={2020 IEEE Int. Conference on Software Maintenance and Evolution (ICSME)},
  pages={569--580},
  year={2020},
  organization={IEEE}
}

@book{newman2021building,
  title={Building microservices},
  author={Newman, Sam},
  year={2021},
  publisher={" O'Reilly Media, Inc."}
}

@article{schulman2017proximal,
  title={Proximal policy optimization algorithms},
  author={Schulman, John and Wolski, Filip and Dhariwal, Prafulla and Radford, Alec and Klimov, Oleg},
  journal={arXiv preprint arXiv:1707.06347},
  year={2017}
}

@inproceedings{van2012kieker,
  title={Kieker: A framework for application performance monitoring and dynamic software analysis},
  author={Van Hoorn, Andr{\'e} and Waller, Jan and Hasselbring, Wilhelm},
  booktitle={Proceedings of the 3rd ACM/SPEC international conference on performance engineering},
  pages={247--248},
  year={2012}
}

@article{hasselbring2020kieker,
  title={Kieker: A monitoring framework for software engineering research},
  author={Hasselbring, Wilhelm and Van Hoorn, Andr{\'e}},
  journal={Software Impacts},
  volume={5},
  pages={100019},
  year={2020},
  publisher={Elsevier}
}

@inproceedings{kiczales1997aspect,
  title={Aspect-oriented programming},
  author={Kiczales, Gregor and Lamping, John and Mendhekar, Anurag and Maeda, Chris and Lopes, Cristina and Loingtier, Jean-Marc and Irwin, John},
  booktitle={European conference on object-oriented programming},
  pages={220--242},
  year={1997},
  organization={Springer}
}

@article{seaman1999qualitative,
  title={Qualitative methods in empirical studies of software engineering},
  author={Seaman, Carolyn B.},
  journal={IEEE Transactions on software engineering},
  volume={25},
  number={4},
  pages={557--572},
  year={1999},
  publisher={IEEE}
}

@article{beck2025tutorial,
  title={A tutorial on meta-reinforcement learning},
  author={Beck, Jacob and Vuorio, Risto and Liu, Evan Zheran and Xiong, Zheng and Zintgraf, Luisa and Finn, Chelsea and Whiteson, Shimon and others},
  journal={Foundations and Trends{\textregistered} in Machine Learning},
  volume={18},
  number={2-3},
  pages={224--384},
  year={2025},
  publisher={Now Publishers, Inc.}
}

@inproceedings{jin2018functionality,
  title={Functionality-oriented microservice extraction based on execution trace clustering},
  author={Jin, Wuxia and Liu, Ting and Zheng, Qinghua and Cui, Di and Cai, Yuanfang},
  booktitle={2018 IEEE Int. Conference on Web Services (ICWS)},
  pages={211--218},
  year={2018},
  organization={IEEE}
}

@inproceedings{kalia2021mono2micro,
  title={Mono2micro: a practical and effective tool for decomposing monolithic java applications to microservices},
  author={Kalia, Anup K and Xiao, Jin and Krishna, Rahul and Sinha, Saurabh and Vukovic, Maja and Banerjee, Debasish},
  booktitle={29th ACM Joint Meeting on European Software Engineering Conference and Symposium on the Foundations of Software Engineering},
  pages={1214--1224},
  year={2021}
}

@inproceedings{kalia2020mono2micro,
  title={Mono2micro: an ai-based toolchain for evolving monolithic enterprise applications to a microservice architecture},
  author={Kalia, Anup K and Xiao, Jin and Lin, Chen and Sinha, Saurabh and Rofrano, John and Vukovic, Maja and Banerjee, Debasish},
  booktitle={Proceedings of the 28th ACM joint meeting on european software engineering conference and symposium on the foundations of software engineering},
  pages={1606--1610},
  year={2020}
}

@article{2022ppoeffectiveness,
  title={The surprising effectiveness of ppo in cooperative multi-agent games},
  author={Yu, Chao and Velu, Akash and Vinitsky, Eugene and Gao, Jiaxuan and Wang, Yu and Bayen, Alexandre and Wu, Yi},
  journal={Advances in neural information processing systems},
  volume={35},
  pages={24611--24624},
  year={2022}
}

@article{al2021microservice,
  title={A microservice decomposition method through using distributed representation of source code},
  author={Al-Debagy, Omar and Martinek, Peter},
  journal={Scalable Computing: Practice and Experience},
  volume={22},
  number={1},
  pages={39--52},
  year={2021}
}

@inproceedings{saidani2019towards,
  title={Towards automated microservices extraction using muti-objective evolutionary search},
  author={Saidani, Islem and Ouni, Ali and Mkaouer, Mohamed Wiem and Saied, Aymen},
  booktitle={Service-Oriented Computing: 17th Int. Conference, ICSOC 2019, Toulouse, France, October 28--31, 2019, Proceedings 17},
  pages={58--6f3},
  year={2019},
  organization={Springer}
}

@ARTICLE{jin2021service,
  author={Jin, Wuxia and Liu, Ting and Cai, Yuanfang and Kazman, Rick and Mo, Ran and Zheng, Qinghua},
  journal={IEEE Transactions on Software Engineering}, 
  title={Service Candidate Identification from Monolithic Systems Based on Execution Traces}, 
  year={2021},
  volume={47},
  number={5},
  pages={987-1007},
  doi={10.1109/TSE.2019.2910531}}

@inproceedings{Francesco2018industrial,
  title = { Migrating towards microservice architectures: an industrial survey },
  author = { Paolo Di Francesco and Patricia Lago and Ivano Malavolta },
  booktitle = { 15th Intl. Conference on Software Architecture (ICSA) },
  pages = { 29--2909 },
  year = { 2018 }
}

@inproceedings{Silva2019,
  year = { 2019 },
  author = { Hugo S. da Silva and Glauco Carneiro and Miguel Monteiro },
  title = { Towards a Roadmap for the Migration of Legacy Software Systems to a Microservice based Architecture },
  booktitle = { 9th Intl. Conference on Cloud Computing and Services Science (CLOSER) },
  pages = { 1--11 }
}

@article{Wang2020,
  year = { 2021 },
  volume = { 26 },
  number = { 4 },
  author = { Yingying Wang and Harshavardhan Kadiyala and Julia Rubin },
  title = { Promises and challenges of microservices: an exploratory study },
  journal = { Empirical Software Engineering },
  pages = { 1--44 }
}

@book{evans2004domain,
  title = { Domain-driven design: tackling complexity in the heart of software },
  author = { Eric Evans },
  year = { 2004 },
  publisher = { Addison-Wesley Professional }
}

@book{millett2015patterns,
  title={Patterns, principles, and practices of domain-driven design},
  author={Millett, Scott and Tune, Nick},
  year={2015},
  publisher={John Wiley \& Sons}
}

@ARTICLE{nsga-iii,
  author={Deb, Kalyanmoy and Jain, Himanshu},
  journal={IEEE Transactions on Evolutionary Computation}, 
  title={An Evolutionary Many-Objective Optimization Algorithm Using Reference-Point-Based Nondominated Sorting Approach, Part I: Solving Problems With Box Constraints}, 
  year={2014},
  volume={18},
  number={4},
  pages={577-601},
  doi={10.1109/TEVC.2013.2281535}}

@inproceedings{laliwala2008event,
  title={Event-driven service-oriented architecture},
  author={Laliwala, Zakir and Chaudhary, Sanjay},
  booktitle={2008 International Conference on Service Systems and Service Management},
  pages={1--6},
  year={2008},
  organization={IEEE}
}

@book{corbin2014basics,
  title = { Basics of qualitative research: Techniques and procedures for developing grounded theory },
  author = { Juliet Corbin and Anselm Strauss },
  year = { 2014 },
  publisher = { Sage publications }
}

@inproceedings{Luz2018,
  author = { Welder Luz and Everton Agilar and Marcos C\'{e}sar de Oliveira and Carlos Eduardo R. de Melo and Gustavo Pinto and Rodrigo Bonif\'{a}cio },
  title = { An Experience Report on the Adoption of Microservices in Three Brazilian Government Institutions },
  year = { 2018 },
  pages = { 32--41 },
  booktitle = { 32nd Brazilian Symposium on Software Engineering (SBES) }
}

@inproceedings{Wolfart2021,
author = {Wolfart, Daniele and Assun\c{c}\~{a}o, Wesley K. G. and da Silva, Ivonei F. and Domingos, Diogo C. P. and Schmeing, Ederson and Villaca, Guilherme L. Donin and Paza, Diogo do N.},
title = {Modernizing Legacy Systems with Microservices: A Roadmap},
year = {2021},
isbn = {9781450390538},
publisher = {ACM},
address = {New York, NY, USA},
url = {https://doi.org/10.1145/3463274.3463334},
doi = {10.1145/3463274.3463334},
booktitle = {Evaluation and Assessment in Software Engineering},
pages = {149–159},
numpages = {11},
location = {Trondheim, Norway},
series = {EASE 2021}
}

@Inbook{Dragoni2017,
author="Dragoni, Nicola
and Giallorenzo, Saverio
and Lafuente, Alberto Lluch
and Mazzara, Manuel
and Montesi, Fabrizio
and Mustafin, Ruslan
and Safina, Larisa",
title="Microservices: Yesterday, Today, and Tomorrow",
bookTitle="Present and Ulterior Software Engineering",
year="2017",
publisher="Springer Int. Publishing",
address="Cham",
pages="195--216",
isbn="978-3-319-67425-4",
doi="10.1007/978-3-319-67425-4_12",
url="https://doi.org/10.1007/978-3-319-67425-4_12"
}

@INPROCEEDINGS {krause2020,
author = {A. Krause and C. Zirkelbach and W. Hasselbring and S. Lenga and D. Kroger},
booktitle = {Int. Conference on Software Architecture Companion (ICSA-C)},
title = {Microservice Decomposition via Static and Dynamic Analysis of the Monolith},
year = {2020},
pages = {9-16},
doi = {10.1109/ICSA-C50368.2020.00011},
publisher = {IEEE},
}

@article{sellami2022,
title = {Improving microservices extraction using evolutionary search},
journal = {Information and Software Technology},
volume = {151},
pages = {106996},
year = {2022},
issn = {0950-5849},
doi = {https://doi.org/10.1016/j.infsof.2022.106996},
url = {https://www.sciencedirect.com/science/article/pii/S0950584922001264},
author = {Khaled Sellami and Ali Ouni and Mohamed Aymen Saied and Salah Bouktif and Mohamed Wiem Mkaouer}
}

@inproceedings{mancoridis1998using,
  title={Using automatic clustering to produce high-level system organizations of source code},
  author={Mancoridis, Spiros and Mitchell, Brian S and Rorres, Chris and Chen, Y and Gansner, Emden R},
  booktitle={Proceedings. 6th International Workshop on Program Comprehension. IWPC'98 (Cat. No. 98TB100242)},
  pages={45--52},
  year={1998},
  organization={IEEE}
}

@inproceedings{khadka2014,
author = {Khadka, Ravi and Batlajery, Belfrit V. and Saeidi, Amir M. and Jansen, Slinger and Hage, Jurriaan},
title = {How Do Professionals Perceive Legacy Systems and Software Modernization?},
year = {2014},
isbn = {9781450327565},
publisher = {ACM},
address = {New York, NY, USA},
doi = {10.1145/2568225.2568318},
booktitle = {36th Int. Conference on Software Engineering},
pages = {36–47},
numpages = {12},
location = {Hyderabad, India},
series = {ICSE 2014}
}

@inproceedings{scanniello2010using,
  title={Using the kleinberg algorithm and vector space model for software system clustering},
  author={Scanniello, Giuseppe and D'Amico, Anna and D'Amico, Carmela and D'Amico, Teodora},
  booktitle={2010 IEEE 18th International Conference on Program Comprehension},
  pages={180--189},
  year={2010},
  organization={IEEE}
}

@inproceedings{fan2017migrating,
  title={Migrating monolithic mobile application to microservice architecture: An experiment report},
  author={Fan, Chen-Yuan and Ma, Shang-Pin},
  booktitle={2017 ieee international conference on ai \& mobile services (aims)},
  pages={109--112},
  year={2017},
  organization={IEEE}
}

@INPROCEEDINGS{regressionTest2018,
  author={Gyori, Alex and Legunsen, Owolabi and Hariri, Farah and Marinov, Darko},
  booktitle={2018 IEEE 29th International Symposium on Software Reliability Engineering (ISSRE)}, 
  title={Evaluating Regression Test Selection Opportunities in a Very Large Open-Source Ecosystem}, 
  year={2018},
  volume={},
  number={},
  pages={112-122},
  doi={10.1109/ISSRE.2018.00022}}

@misc{jpetstore_doc,
author = {Wilson Mar},
title = {JPetstore Documentation},
url={https://wilsonmar.github.io/jpetstore/},
note         = {Accessed: 2025-09-11},
year = {2017}
}

@misc{petclinic_doc,
author = {The Spring Petclinic Community},
title = {Petclinic Documentation},
url={https://spring-petclinic.github.io/docs/resources.html},
note         = {Accessed: 2025-09-11},
year = {2020}
}

@misc{edition_doc,
author = {Ldod Archive},
title = {Digital Archive Social Edition Documentation},
url={https://ldod.uc.pt/about/archive},
note         = {Accessed: 2025-09-11},
year = {2023}
}

@misc{roller_doc,
author = {Apache Roller},
title = {Roller Documentation},
url={https://github.com/apache/roller/tree/roller-6.0.x/docs},
note         = {Accessed: 2025-09-11},
year= {2020}
}

@misc{roller_codebase,
author = {Apache},
title = {Apache Roller Code-base},
url={https://github.com/apache/roller},
note = {Accessed: 2025-09-11},
year = {2020}
}

@misc{jpetstore_codebase,
author = {mybatis},
title = {Jpetstore-6 Code-base},
url={https://github.com/mybatis/jpetstore-6},
note = {Accessed: 2025-09-11},
year = {2022}
}

@misc{petclinic_codebase,
  author       = {spring-petclinic},
  title        = {Spring Petclinic Code-base},
  howpublished = {\url{https://github.com/spring-petclinic/spring-framework-petclinic/releases/tag/v5.3.22}},
  note         = {Accessed: 2025-09-11},
year= {2022}
}

@misc{edition_codebase,
author = {social software},
title = {Social Software Edition Code-base},
url={https://github.com/socialsoftware/edition/tree/modular-monolith},
note = {Accessed: 2025-09-11},
year = {2021}
}

@misc{kieker_codebase,
author = {Kieker},
title = {Example Visualization – Assembly Operation Dependency Graph from Kieker.WebGUI},
url={https://kieker-monitoring.net/architecture-discovery/attachment/webgui_assemblyoperationdependencygraph/},
note = {Accessed: 2025-09-11},
year = {2013}
}

@misc{kiekerv15,
author = {Kieker Monitoring},
title = {Kieker Version 1.15 Release},
url={{https://github.com/kieker-monitoring/kieker/releases/tag/1.15}},
note = {Accessed: 2025-09-11},
year = {2021}
}

@misc{kieker_trace,
author = {Kieker Trace-Analysis},
title = {Using Kieker Trace-Analysis},
url={{https://kieker-monitoring.readthedocs.io/en/latest/getting-started/Using-Kieker-Trace-Analysis.html}},
note = {Accessed: 2025-09-11},
year = {2020}
}

@misc{java_parser,
author = {JavaParser},
title = {Static Java Parser},
url={https://github.com/javaparser/javaparser},
note = {Accessed: 2025-09-11},
year = {2021}
}

@article{van2009continuous,
  title={Continuous monitoring of software services: Design and application of the Kieker framework},
  author={van Hoorn, Andr{\'e} and Hasselbring, Wilhelm and Waller, Jan and Ehlers, Jens and Frey, S{\"o}ren and Kieselhorst, Dennis},
  year={2009}
}

@article{Martinez2025,
title = {Migration of monolithic systems to microservices: A systematic mapping study},
journal = {Information and Software Technology},
volume = {177},
pages = {107590},
year = {2025},
issn = {0950-5849},
doi = {https://doi.org/10.1016/j.infsof.2024.107590},
author = {Ana {Martínez Saucedo} and Guillermo Rodríguez and Fabio {Gomes Rocha} and Rodrigo Pereira dos Santos}
}

@ARTICLE{Abgaz2023,
author={Abgaz, Yalemisew and McCarren, Andrew and Elger, Peter and Solan, David and Lapuz, Neil and Bivol, Marin and Jackson, Glenn and Yilmaz, Murat and Buckley, Jim and Clarke, Paul},
journal={IEEE Transactions on Software Engineering}, 
title={Decomposition of Monolith Applications Into Microservices Architectures: A Systematic Review}, 
year={2023},
volume={49},
number={8},
pages={4213-4242},
doi={10.1109/TSE.2023.3287297}
}

@article{Dehghani2022,
  title={Facilitating the migration to the microservice architecture via model-driven reverse engineering and reinforcement learning},
  author={Dehghani, MohammadHadi and Kolahdouz-Rahimi, Shekoufeh and Tisi, Massimo and Tamzalit, Dalila},
  journal={Software and Systems Modeling},
  volume={21},
  number={3},
  pages={1115--1133},
  year={2022},
  publisher={Springer}
}

@article{oumoussa2024evolution,
  title={Evolution of microservices identification in monolith decomposition: A systematic review},
  author={Oumoussa, Idris and Saidi, Rajaa},
  journal={IEEE Access},
  volume={12},
  pages={23389--23405},
  year={2024},
  publisher={IEEE}
}

@article{sellami2025rldec,
  title={Extracting microservices from monolithic systems using deep reinforcement learning},
  author={Sellami, Khaled and Saied, Mohamed Aymen},
  journal={Empirical Software Engineering},
  volume={30},
  number={1},
  pages={1},
  year={2025},
  publisher={Springer}
}

@misc{martin2014srp,
  author       = {Robert C. Martin},
  title        = {The Single Responsibility Principle},
  year         = {2014},
  howpublished = {\url{https://blog.cleancoder.com/uncle-bob/2014/05/08/SingleReponsibilityPrinciple.html}},
  note         = {Accessed: 2025-08-25}
}

@inproceedings{wang2024microservice,
  title={Microservice decomposition techniques: An independent tool comparison},
  author={Wang, Yingying and Bornais, Sarah and Rubin, Julia},
  booktitle={Proceedings of the 39th IEEE/ACM International Conference on Automated Software Engineering},
  pages={1295--1307},
  year={2024}
}

@INPROCEEDINGS{toMicroservice2021,
author={Assunção, Wesley K. G. and Colanzi, Thelma Elita and Carvalho, Luiz and Pereira, Juliana Alves and Garcia, Alessandro and de Lima, Maria Julia and Lucena, Carlos},
booktitle={IEEE Int. Conference on Software Analysis, Evolution and Reengineering (SANER)}, 
title={A Multi-Criteria Strategy for Redesigning Legacy Features as Microservices: An Industrial Case Study}, 
year={2021},
pages={377-387},
doi={10.1109/SANER50967.2021.00042}
}

@book{gradecki2003mastering,
  title={Mastering AspectJ: aspect-oriented programming in Java},
  author={Gradecki, Joseph D and Lesiecki, Nicholas},
  year={2003},
  publisher={John Wiley \& Sons}
}

@article{Tuli2014,
author = {Tuli, Anupriya and Hasteer, Nitasha and Sharma, Megha and Bansal, Abhay},
title = {Empirical investigation of agile software development: cloud perspective},
year = {2014},
issue_date = {July 2014},
publisher = {ACM},
address = {New York, NY, USA},
volume = {39},
number = {4},
issn = {0163-5948},
doi = {10.1145/2632434.2632447},
journal = {SIGSOFT Softw. Eng. Notes},
month = {aug},
pages = {1–6},
numpages = {6}
}

@inproceedings{Tizzei2017,
author = {Tizzei, Leonardo P. and Nery, Marcelo and Segura, Vin\'{\i}cius C. V. B. and Cerqueira, Renato},
title = {Using Microservices and Software Product Line Engineering to Support Reuse of Evolving Multi-tenant \mbox{SaaS}},
booktitle = {21st Int. Systems and Software Product Line Conference (SPLC)},
year = {2017},
isbn = {978-1-4503-5221-5},
location = {Sevilla, Spain},
pages = {205--214},
acmid = {3106224},
}

@article{mcburney2016empirical,
  title={An empirical study of the textual similarity between source code and source code summaries},
  author={McBurney, Paul W and McMillan, Collin},
  journal={Empirical Software Engineering},
  volume={21},
  number={1},
  pages={17--42},
  year={2016},
  publisher={Springer}
}

\appendix

\end{document}